\documentclass[twocolumn]{aastex7}

\begin{document}

\title{Exploring RR Lyrae Variable Stars in the Vera C. Rubin Observatory Data Preview 1}

\author[0000-0001-8771-7554,gname=Chow-Choong, sname=Ngeow]{Chow-Choong Ngeow}
\affil{Graduate Institute of Astronomy, National Central University, 300 Jhongda Road, 32001 Jhongli, Taiwan}
\affil{Taiwan Astronomical Research Alliance (TARA)}
\email[show]{cngeow@astro.ncu.edu.tw}

\author[0000-0001-6147-3360,gname=Anupam, sname=Bhardwaj]{Anupam Bhardwaj}
\affil{Inter-University Center for Astronomy and Astrophysics (IUCAA), Post Bag 4, Ganeshkhind, Pune 411 007, India}
\email{anupam.bhardwajj@gmail.com}


\begin{abstract}

We investigate the properties of known RR Lyrae in the Vera C. Rubin Observatory Data Preview 1 (DP1)  fields and compare those with the predictions based on stellar pulsation models tailored to the Legacy Survey of Space and Time (LSST) filters. The cross-match of the DP1 data with two public variable star catalogs resulted in $\sim 600$ RR Lyrae with adequate light curve sampling in five (out of seven) DP1 fields. The majority of RR Lyrae are in the 47 Tucanae and Fornax fields. We estimated photometric metallicities for these RR Lyrae using the theoretical metallicity-color relation based on $gri$-band data, and find a good agreement with literature values where the light curve sampling is sufficient for fitting template light curves accurately. The distance modulus to all RR Lyrae in DP1 fields were determined using the theoretical period-luminosity-metallicity (PLZ) relations and the $W_{gr}$ period-Wesenheit-metallicity (PWZ) relation which has the smallest metallicity term. The distances based on PWZ relations are in good agreement with the literature values with a mean offset of $0.01\pm0.36$~mag. However, the PLZ-based distance moduli are systematically large which could be due to the theoretical calibration uncertainties that include evolved horizontal branch models. The predicted period-amplitude relations based on evolved models are also inconsistent with the amplitudes based on DP1 light curves. We conclude that the metallicity and distance estimates are sensitive to the template fitting to sparsely sampled light curves in DP1 data and future data release will significantly improve these determinations for RR Lyrae stars.
  
\end{abstract}

\section{Introduction}

Prior to the beginning of the full Legacy Survey of Space and Time (LSST), the NSF-DOE Vera C. Rubin Observatory, funded by the U.S. National Science Foundation and the U.S. Department of Energy's Office of Science, has released a set of commissioning data on 30 June, 2025. The released data, known as the Data Preview 1 \citep[hereafter DP1,][]{10.71929/rubin/2570308}, was collected using the Rubin Commissioning Camera \citep[the LSSTComCam,][]{lsstcomcam} on 48 nights in late 2024. DP1 focused on obtaining time-series data on seven pre-selected fields, each with a size of $\sim1.3^\circ \times 1.3^\circ$. More details regarding DP1 can be found in the accompanying release paper \citep{RTN-095}.

RR Lyrae are old-population pulsating stars that are well-known distance indicators \citep[e.g., reviews by][]{beaton2018,bhardwaj2020}. These stars are excellent population tracers to map the sub-structures of the Galactic halo \citep[see][and reference therein]{hernitschek2018,stringer2021,wang2022,feng2024} or constrain the distances to the dwarf galaxies \citep[for examples, see][]{monelli2022,nagarajan2022,bhardwaj2024,ngeow2026}. Indeed, RR Lyrae are one of the science targets for LSST \citep{ivezic2019,hambleton2023}. After the release of DP1, \citet{malanchev2025} demonstrated that several known RR Lyrae can be recovered from the DP1 light-curve data. Similarly, \citet{choi2025}, who focused their investigation on the most crowded DP1 field -- the 47Tuc (47 Tucanae) field, recovered two known RR Lyrae from the {\tt Object} catalog and more variables using the {\tt DiaObject} catalog of the DP1 data.\footnote{Schema for these tables can be found in \url{https://sdm-schemas.lsst.io/dp1.html}\label{fn5}}

The aforementioned studies only demonstrated that the sparse DP1 light curves could be recovered for a handful of known RR Lyrae in some of the DP1 fields. The main goal of this work is to explore and study the known RR Lyrae stars in detail in all of the seven DP1 fields (if available). In particular, we compared the empirical multiband light curve amplitudes and colors with those predicted from the theoretical models \citep{marconi2022}. We also derived the metallicity and hence the distances to RR Lyrae using the period-luminosity-metallicity (PLZ) and/or period-Wesenheit-metallicity (PWZ) relations presented in \citet{marconi2022}. The pulsation models and theoretical relations investigated in \citet{marconi2022} were specifically targeted to the Rubin LSST filters, facilitating a first direct comparison between models and real DP1 data for RR Lyrae.

\section{Known RR Lyrae in the DP1 Fields}  \label{sec_rrl}

There are several data products released together with DP1 \citep[for more details, see][]{RTN-095}, including catalogs storing time-dependent measurements based on the single-visit images and the difference imaging analysis (Dia) pipeline. The LSDB \citep[Large Scale Database,][]{lsdb2025}\footnote{Available at \url{https://lsdb.io/}} framework has transformed the DP1 catalogs to two science-ready catalogs, the {\tt Object} catalog and the {\tt DiaObject} catalog. Note that these two catalogs were used in \citet{choi2025} for crossmatching the known RR Lyrae in the 47Tuc field. An additional advantage of using LSDB is the International Variable Star Index \citep[VSX,][]{watson2006} database was also included in LSDB, making it a straight-forward to search and crossmatch the known RR Lyrae stars in all of the DP1 fields.

Following the instructions in LSDB, we crossmatched known RR Lyrae in VSX database to both of the {\tt Object} catalog and the {\tt DiaObject} catalog using a search radius of $1\arcsec$. The numbers of crossmatched RR Lyrae in each DP1 fields are summarized in Table \ref{tab1}, and referred as the VSX sample. Even though the ECDFS field contains the most intense time-series observations among the DP1 fields, unfortunately it did not contain any known RR Lyrae. Other four low to moderate crowded fields (EDFS, Rubin95, Rubin38, and Seagull) have the same number of RR Lyrae in both catalogs. The last two crowded fields, Fornax and 47Tuc, have more RR Lyrae retrieved from the {\tt DiaObject} catalog, due to the reasons extensively discussed in \citet{choi2025}. We emphasize that the RR Lyrae in the 47Tuc field are mostly associated to the Small Magellanic Cloud, and do not belong to the globular cluster 47 Tucanae. Note that due to extreme crowded environment, the central parts of both fields were not included in the DP1. The different numbers of RR Lyrae in 47Tuc field retrieved in Table \ref{tab1} and in \citet{choi2025} is mainly caused by the use of different variable star catalogs.  

\begin{deluxetable}{lrrrc}
  \label{tab1}
  \tabletypesize{\scriptsize}
  \tablecaption{Number of known RR Lyrae in the DP1 fields.}
  \tablewidth{10pt}
  \tablehead{
    \colhead{DP1 Field\tablenotemark{a}} &
    \colhead{$N_{\mathrm{ep}}$\tablenotemark{b}} &
    \colhead{$N_{OBJ}$\tablenotemark{c}} &
    \colhead{$N_{DIA}$\tablenotemark{c}} &
    \colhead{$N_{F}(OBJ/DIA)$\tablenotemark{d}} 
  }
  \startdata
  \multicolumn{5}{c}{VSX sample} \\
  47Tuc    &  4 & 25  & 62 &  1/36  \\  
  ECDFS    & 21 &  0  &  0 &  0/0 \\  
  EDFS     &  9 & 1   &  1 &  1/1 \\ 
  Fornax   &  2 & 3   & 10 &  0/7 \\   
  Rubin95  & 10 & 2   &  2 &  2/2 \\  
  Rubin38  &  5 & 4   &  4 &  1/1 \\ 
  Seagull  &  4 & 1   &  1 &  0/0 \\ 
  \multicolumn{5}{c}{DES sample} \\
  Fornax   & $\cdots$ & 393   & 617  & 146/517 \\   
  \enddata 
  \tablenotetext{a}{Abbreviated DP1 field names following the order given in the Table 1 of \citet{RTN-095}.}
  \tablenotetext{b}{Number of observed epochs (or nights) for each fields, adopted from \url{https://dp1.lsst.io/overview/observations.html}.}
  \tablenotetext{c}{Number of the known RR Lyrae after crossmatching to the {\tt Object} and {\tt DiaObject} catalogs in LSDB, respectively.}
  \tablenotetext{d}{Final number of the RR Lyrae in the samples.}
\end{deluxetable}

The number of known RR Lyrae in the Fornax field is surprisingly low, because the Fornax dwarf spheroidal (dSph) galaxy is known to host a large population of RR Lyrae \citep{bersier2002,fiorentino2017,braga2022}. This implies the VSX database in this field is still incomplete. To remedy this, we crossmatched the LSDB catalogs with a RR Lyrae catalog available from \citet{stringer2021}, because this catalog fully covered the Fornax field. As a result, the numbers of known RR Lyrae were increased to 393 and 617 when crossmatched to the {\tt Object} and the {\tt DiaObject} catalogs, respectively. We refer to this sample of RR Lyrae as the DES sample because the data used in \citet{stringer2021} was from the Dark Energy Survey \citep[DES,][]{des2016}.

\begin{figure*}
  \epsscale{1.1}
  \plottwo{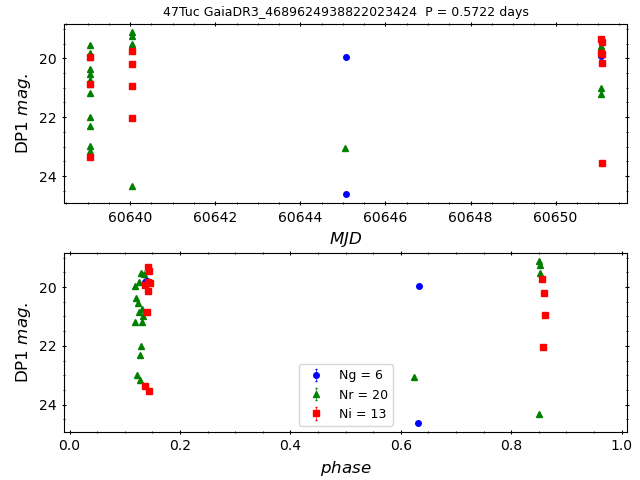}{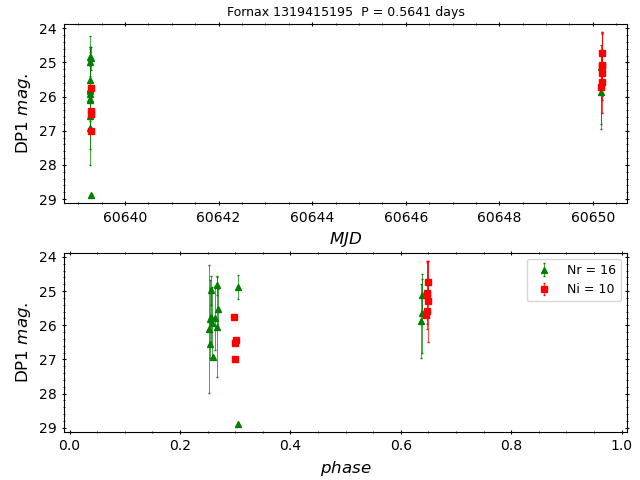}
  \caption{Examples of two RR Lyrae, one from the 47Tuc and Fornax fields, respectively, with light curves exhibiting a large scatter at a given epoch. Both of these RR Lyrae only have two or four epochs of data available in the DP1. Upper and lower panels present multi-band light curves in time and in pulsation phases (after folded with published period), respectively.}
  \label{fig_badlc1}
\end{figure*}

For the aforementioned two samples of RR Lyrae, we adopted their published periods ($P$) in VSX or \citet{stringer2021} catalogs. In addition, we have also obtained their $E(B-V)$ values from the \citet{schlegel1998} dust maps by using the the {\tt dustmaps} \citep{green2018} code. Values of the extinction coefficients in LSST filters, $R_\lambda$ where $\lambda=\{u,\ g,\ r,\ i,\ z,\ y\}$, were adopted from \citet{schlafly2011} with $R_V=3.1$.

\section{DP1 Light-Curves} \label{sec_lc}

For the crossmatched RR Lyrae, we extracted two sets of light curves via LSDB. The first set of light curves are those nested in the {\tt Object} catalog, with the {\tt objectForcedSource} attribute. In case of the {\tt DiaObject} catalog, two types of nested light curves are available: the {\tt diaSource} light curves (which are based on the detected sources on the Dia images) and the {\tt diaObjectForcedSource} light curves (which are based on the force photometry for the Dia sources on the Dia images). Following \citet{choi2025}, we adopted the {\tt diaObjectForcedSource} light curves as our second set of light curves, because in general they contain more data-points than the {\tt diaSource} light curves. All of the PSF-based flux measurements in these light curves have been converted to the magnitudes within the LSDB framework \citep{malanchev2025}.

Since not all of the light curves are science ready, we applied a variety of criteria to filter out unsuitable data points. For each RR Lyrae light curve in each filter, we first excluded data points with any of the following flags set to ``True'':  pixelFlags\_nodata, pixelFlags\_bad, pixelFlags\_cr, pixelFlags\_edge, pixelFlags\_saturated and psfFlux\_flag. These flagged data points tend to be outliers in the light curves. A short description of these flags can be found in the schema (see footnote $\ref{fn5}$) of DP1 database. Further, we only kept those light curves that have more than five data points after excluding the flagged data. We have also eliminated a small number of light curves that exhibit unusually large scatter at a given epoch, with two examples shown in Figure \ref{fig_badlc1}. These steps removed the $g$-band and the $y$-band light curves for RR Lyrae in the Fornax field and the 47Tuc field, respectively, as well as the only RR Lyrae in the Seagull field. 

After applying the aforementioned criteria, there are 1, 2, 1, and 0 RR Lyrae left in the EDFS, Rubin95, Rubin38, and Seagull field, respectively (see the last column of Table \ref{tab1}). DP1 light curves for these four RR Lyrae were analyzied in Section \ref{sec_3field}. For RR Lyrae in the 47Tuc and the Fornax field, their sparse DP1 light curves required an additional treatment, which is discussed further in Section \ref{sec_2field}.

\subsection{Template Light Curves Fitting} \label{sec_tmplc}

Recently, \citet{braga2024} published a set of $griz$-band RR Lyrae template light curves to be applicable to Rubin LSST data. Using the accompanying python codes,\footnote{The codes are available at \url{https://github.com/vfbraga/RRL_lcvtemplate_griz_LSST}} we fit these template light curves to the VSX and DES samples. These templates provide a mean magnitude (and its associated error) based on the best-fit template light curve in a given filter, as well as the amplitude $A_\lambda$. For the $u$- and $y$-band light curves, we adopted the $g$- and $z$-band template light curves, respectively, assuming light curves in these two bands resemble those in the $u$- and $y$-band, respectively. Note that the $z$-band template light curves were only available for the fundamental mode RRab stars \citep{braga2024}, hence we did not fit the $zy$-band light curves to the only first-overtone RRc star in the Rubin95 field.

The template light curve fitting codes provided by \citet{braga2024} take the pulsation period $P$ as a mandatory input parameter, together with the amplitude $A_\lambda$ and reference epoch $t_0$ as optional input parameters. For those RR Lyrae which have light curves data at least in two filters, we fixed the reference epoch $t_0$ (i.e. at the time of maximum light) from the filter which has the most number of data points, usually it is the $r$-band. This is because the maximum light of multiband light curves should occur nearly at the same phase with possible variations due the phase lag between different filters. In some cases, $t_0$ in a given filter was adjusted such that the maximum light phases are well aligned in different filters. 

\begin{figure*}
  \epsscale{1.1}
    \centering
    \begin{tabular}{ccc}
      \includegraphics[width=5.75cm]{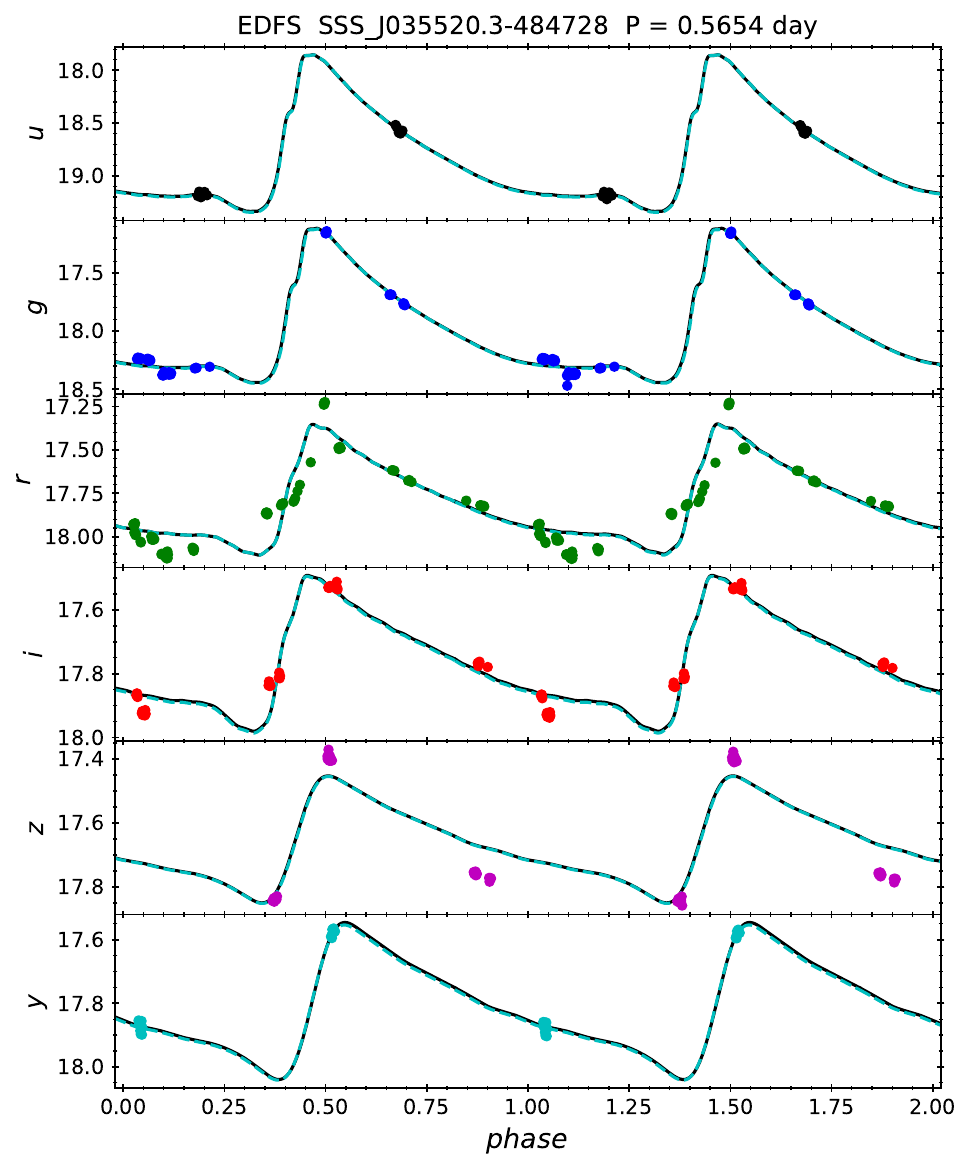} & \includegraphics[width=5.75cm]{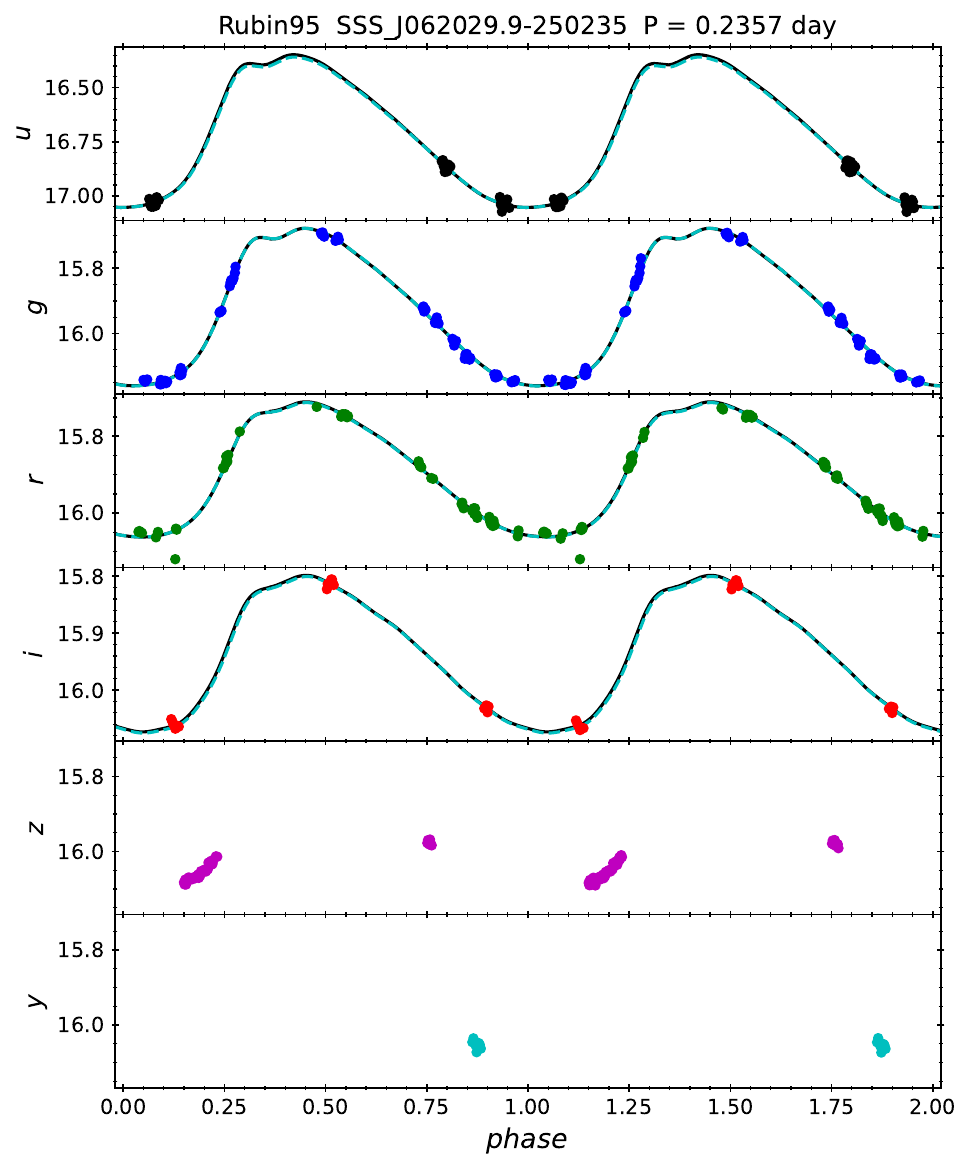} & \includegraphics[width=5.75cm]{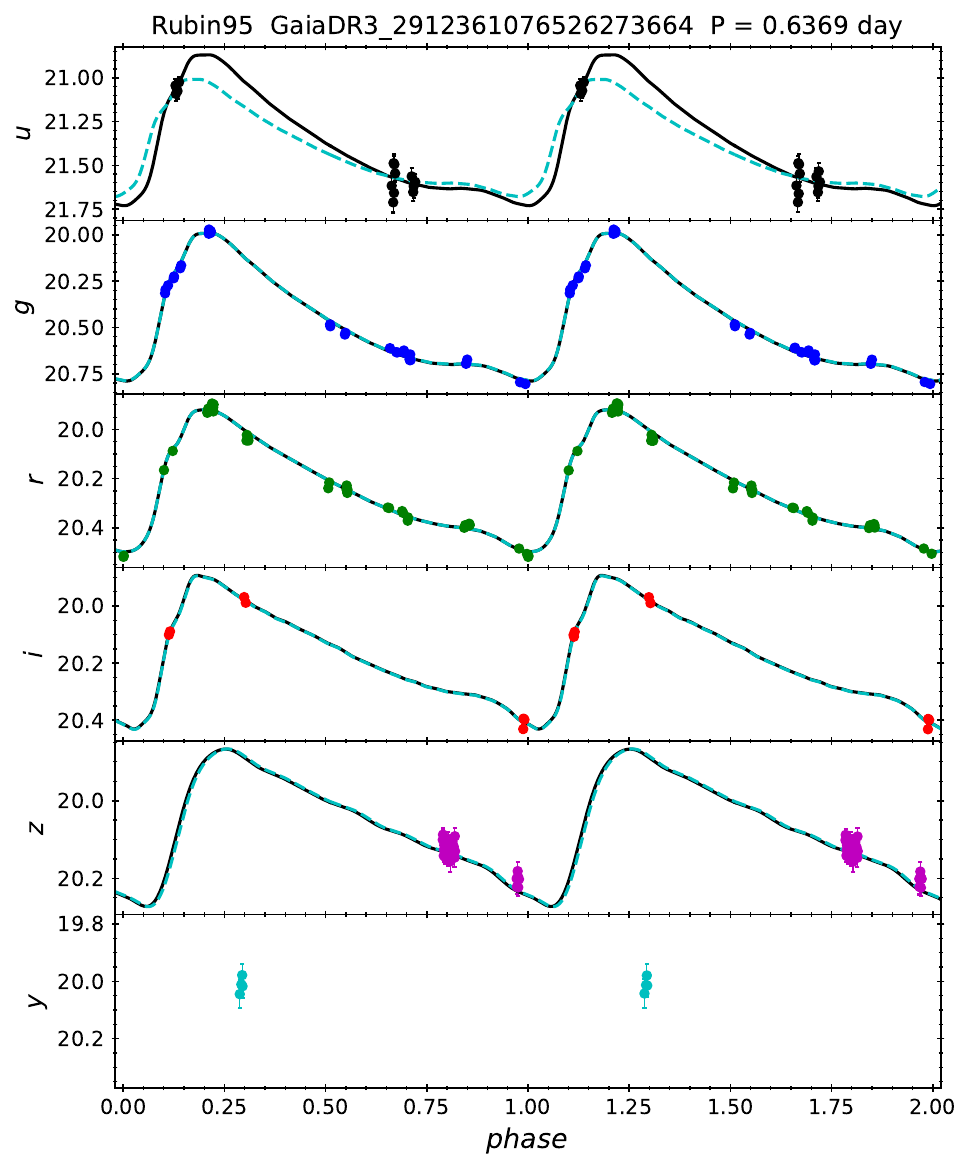} \\
    \end{tabular}
    \caption{The multiband light curves for the three known RR Lyrae in the EDFS field and the Rubin95 field. The best-fit template light curves to the {\tt objectForcedSource} and the {\tt diaObjectForcedSource} data are shown as black solid curves and cyan dashed curves, respectively. We did not fit the template light curves to the $zy$-band data for the RRc star in the Rubin95 field (the middle panels) because \citet{braga2024} did not derive the $z$-band template light curves for RRc stars. On the other hand, fitting the $z$-band template light curve to the $y$-band data for the RRab star in the Rubin95 field (the lowest-right panel) gives a spurious large error ($>5$~mag) on the mean magnitude. Hence we excluded this $y$-band data in template light curve fitting.}
  \label{fig_lc1}
\end{figure*}

\begin{figure*}
  \epsscale{1.1}
  \centering
  \begin{tabular}{ccc}
    \includegraphics[width=5.75cm]{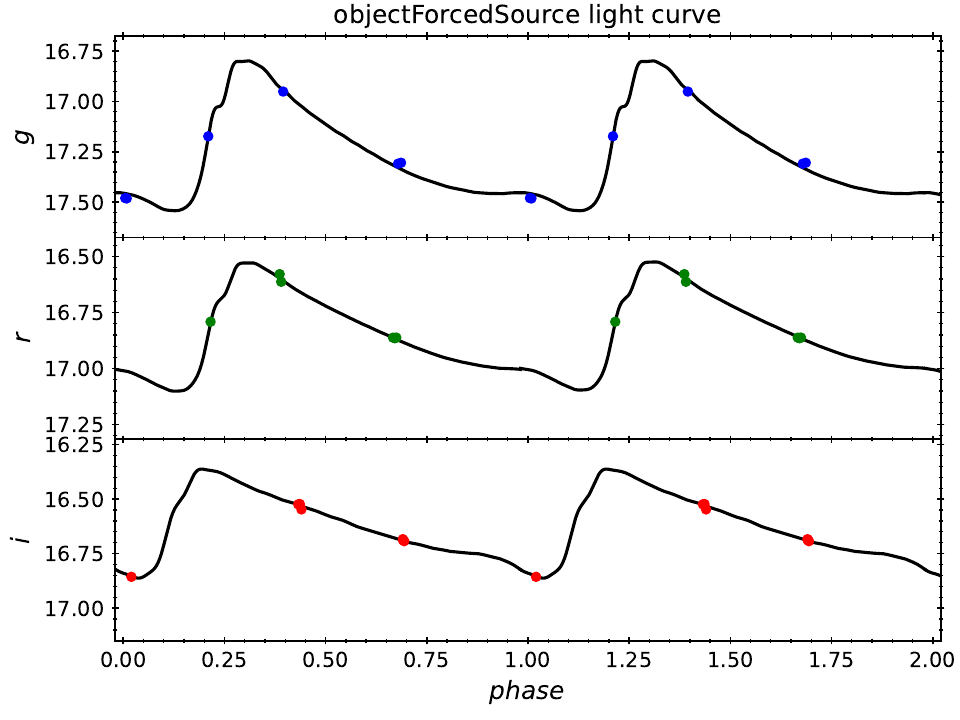} & \includegraphics[width=5.75cm]{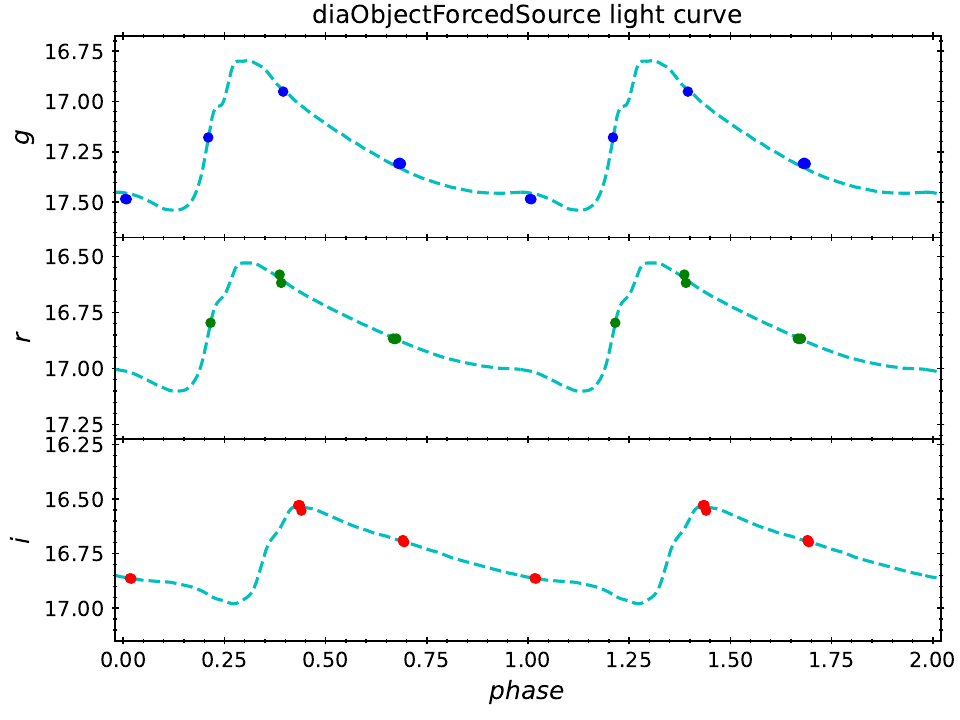} & \includegraphics[width=5.75cm]{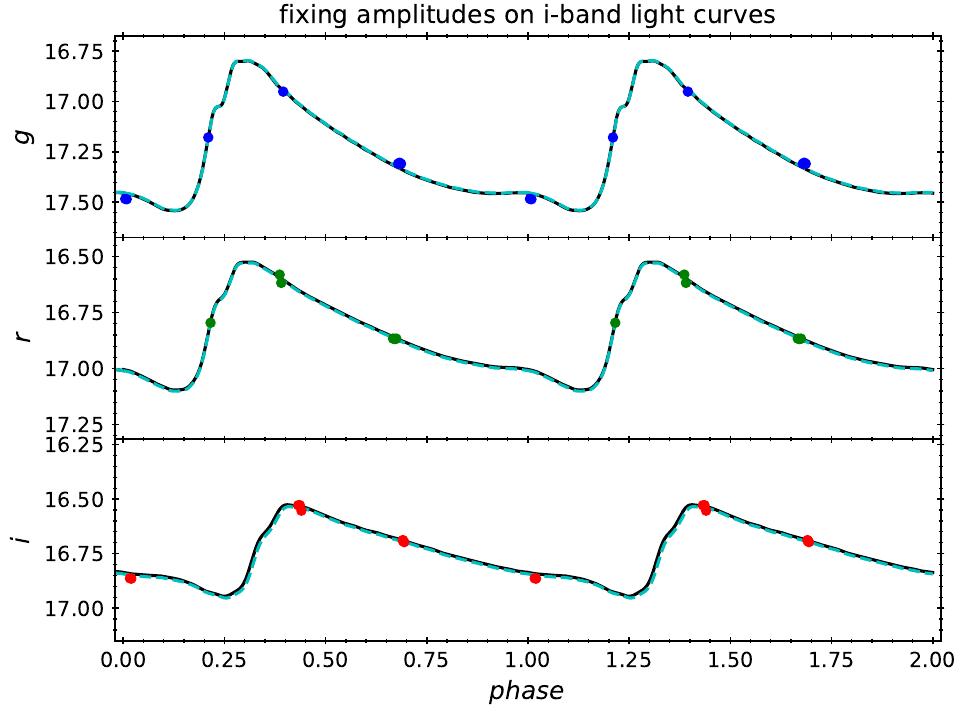} \\
  \end{tabular}
  \caption{Comparison of the best-fit $gri$-band template light curves for CSS J023424.4+072337 in the Rubin38 field. The left and middle panels show the {\tt objectForcedSource} and the {\tt diaObjectForcedSource} light curves, respectively, together with the best fit template light curves. In the right panels, both sets of best-fit template light curves were plotted together, along with the improved $i$-band template light curves by fixing the input amplitudes (see text for details).}
  \label{fig_lc38}
\end{figure*}

\begin{figure}
  \epsscale{1.1}
  \plotone{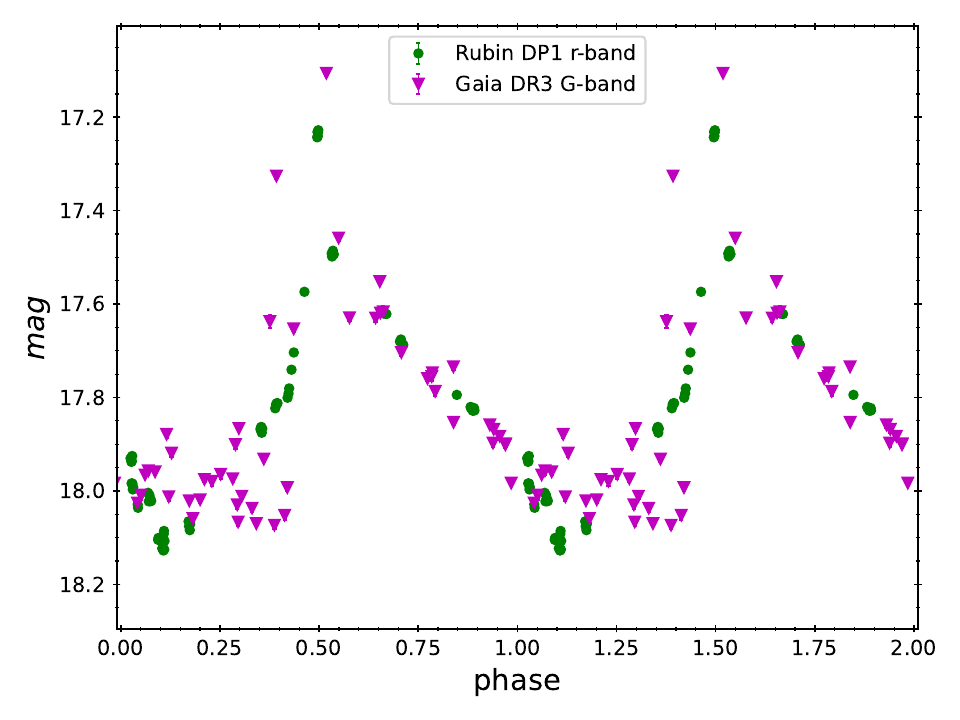}
  \caption{Comparison of the DP1 $r$-band and the Gaia DR3 $G$-band light curves for the only known RR Lyrae, SSS J035520.3-484728, in the EDFS field.}
  \label{fig_edfs}
\end{figure}

The numbers of data points on the remaining LSDB-extracted light curves vary from 6 to $\sim70$ (on the $r$-band light curves of the only RR Lyrae in the EDFS field). We note that an epoch of DP1 observation consists of 5 to 20 visits, each with a 30~second exposure time in a given filter \citep[][except in the $u$-band that has a slightly longer exposure time]{RTN-095}. Hence, some light curve data points, after folded with the pulsation periods, could cluster on specific phases representing the epoch of observations.
Some examples of such light curves containing numerous data points can be seen in Figure \ref{fig_badlc1}, or on the $y$-band light curves shown in Figure \ref{fig_lc1}. This has an impact on the light curve analysis of RR Lyrae located in the 47Tuc and Fornax field. The numbers of epoch observed in each of the DP1 fields are listed in the second column of Table \ref{tab1}.

\begin{deluxetable*}{lcccc}
  \label{tab_vsx}
  \tabletypesize{\scriptsize}
  \tablecaption{Summary of results for the known RR Lyrae in the EDFS, Rubin95, and Rubin38 field.}
  \tablewidth{10pt}
  \tablehead{
    \colhead{} &
    \colhead{EDFS Field } &
    \colhead{Rubin95 Field} &
    \colhead{Rubin95 Field} &
    \colhead{Rubin38 Field} 
  }
  \startdata
  VSX Name     & SSS J035520.3-484728  & SSS J062029.9-250235  &  GaiaDR3 2912361076526273664  & CSS J023424.4+072337 \\
  Mode         & RRab   & RRc & RRab  & RRab \\
  $P$ (days)& 0.56536    & 0.23567 & 0.63694  & 0.61965 \\
  $E(B-V)$ (mag) & 0.006  & 0.042 & 0.041  & 0.178 \\
  $\mathrm{[Fe/H]}_{\mathrm{lit}}$ (dex) & $-2.13\pm0.07$\tablenotemark{a} & $-0.82\pm0.20$\tablenotemark{b} & $-2.26\pm0.05$\tablenotemark{c} & $-2.33\pm0.41$\tablenotemark{b} \\
  $\mu_{\mathrm{lit}}$ (mag)&  $17.19$\tablenotemark{a} & $14.86\pm0.20$\tablenotemark{b} & $\cdots$ & $16.16\pm0.24$\tablenotemark{b} \\
  \hline
  \multicolumn{5}{c}{Using {\tt objectForcedSource} light curves} \\
  \hline
  DP1 objID & 592913981740430064  & 614436990573616692 & 614431011979142887  & 648375409829750839 \\
  $N_u,N_g,N_r,N_i,N_z,N_y$ & 12,51,72,37,36,19 & 24,68,54,19,38,8 & 15,31,38,7,32,4 & $\cdots$,6,7,7,$\cdots$,$\cdots$ \\
  $\langle u \rangle$ (mag)  & $18.821\pm0.008$  & $16.711\pm0.004$ & $21.386\pm0.066$  & $\cdots$ \\
  $\langle g \rangle$ (mag)  & $17.981\pm0.001$  & $15.925\pm0.001$ & $20.470\pm0.002$  & $17.260\pm0.001$ \\
  $\langle r \rangle$ (mag)  & $17.802\pm0.001$  & $15.890\pm0.001$ & $20.239\pm0.001$  & $16.848\pm0.001$ \\
  $\langle i \rangle$ (mag)  & $17.766\pm0.001$  & $15.938\pm0.001$ & $20.175\pm0.005$  & $16.625\pm0.003$ \\
  $\langle z \rangle$ (mag)  & $17.732\pm0.012$  & $\cdots$         & $20.063\pm0.013$  & $\cdots$ \\
  $\langle y \rangle$ (mag)  & $17.810\pm0.003$  & $\cdots$         & $\cdots$  & $\cdots$ \\
  $A_u,\ A_g,\ A_r$ (mag)& $1.486,\ 1.327,\ 0.750$  & $0.706,\ 0.483,\ 0.352$ & $0.862,\ 0.800,\ 0.579$ & $\cdots,\ 0.744,\ 0.572$ \\
  $A_i,\ A_z,\ A_y$ (mag)& $0.489,\ 0.680,\ 0.495$  & $0.275,\ \cdots,\ \cdots$  & $0.538,\ 0.405,\ \cdots$         & $0.501,\ \cdots,\ \cdots$ \\
  $\mathrm{[Fe/H]}_{gri}$ (dex)& $-1.86\pm0.33$ & $-1.79\pm0.33$ & $-1.73\pm0.38$  & $-3.59\pm0.35$  \\
  $\mu_W$ (mag)             & $17.226\pm0.073$ & $15.118\pm0.062$ & $19.648\pm0.074$ & $15.807\pm0.077$ \\
  $\mu_g$ (mag)             & $17.768\pm0.228$ & $15.581\pm0.227$ & $20.115\pm0.231$ & $16.892\pm0.237$ \\
  $\mu_r$ (mag)             & $17.506\pm0.144$ & $15.209\pm0.104$ & $19.913\pm0.146$ & $16.500\pm0.151$ \\
  $\mu_i$ (mag)             & $17.452\pm0.118$ & $15.180\pm0.086$ & $19.865\pm0.121$ & $16.369\pm0.123$ \\
  $\mu_z$ (mag)             & $17.377\pm0.109$ & $\cdots$         & $19.730\pm0.112$ & $\cdots$ \\
  $\mu_y$ (mag)             & $17.451\pm0.124$ & $\cdots$         & $\cdots$         & $\cdots$ \\ 
  \hline
  \multicolumn{5}{c}{Using {\tt diaObjectForcedSource} light curves} \\
  \hline
  DP1 diaID & 592913981740417050  & 614436990573609020 & 614431011979133047   & 648375409829740567 \\
  $N_u,N_g,N_r,N_i,N_z,N_y$ & 13,54,73,37,38,19 & 29,72,64,21,52,8 & 16,32,39,8,32,4 & $\cdots$,7,7,8,$\cdots$,$\cdots$ \\
  $\langle u \rangle$ (mag)  & $18.827\pm0.029$  & $16.716\pm0.004$ & $21.410\pm0.026$  & $\cdots$ \\
  $\langle g \rangle$ (mag)  & $17.987\pm0.001$  & $15.926\pm0.001$ & $20.469\pm0.002$  & $17.259\pm0.001$ \\
  $\langle r \rangle$ (mag)  & $17.806\pm0.001$  & $15.891\pm0.001$ & $20.239\pm0.001$  & $16.853\pm0.001$ \\
  $\langle i \rangle$ (mag)  & $17.770\pm0.001$  & $15.940\pm0.001$ & $20.176\pm0.005$  & $16.768\pm0.002$ \\
  $\langle z \rangle$ (mag)  & $17.734\pm0.012$  & $\cdots$         & $20.063\pm0.024$  & $\cdots$ \\
  $\langle y \rangle$ (mag)  & $17.815\pm0.003$  & $\cdots$         & $\cdots$  & $\cdots$ \\
  $A_u,\ A_g,\ A_r$ (mag)& $1.494,\ 1.327,\ 0.754$  & $0.695,\ 0.484,\ 0.350$ & $0.669,\ 0.798,\ 0.579$  & $\cdots,\ 0.743,\ 0.574$ \\
  $A_i,\ A_z,\ A_y$ (mag)& $0.492,\ 0.676,\ 0.489$  & $0.275,\ \cdots,\ \cdots$    & $0.538,\ 0.406,\ \cdots$          & $0.447,\ \cdots,\ \cdots$ \\
  $\mathrm{[Fe/H]}_{gri}$ (dex)& $-1.82\pm0.33$  & $-1.78\pm0.33$ & $-1.70\pm0.38$  & $+1.41\pm0.34$ \\
  $\mu_W$ (mag)             & $17.224\pm0.073$ & $15.122\pm0.062$ & $19.650\pm0.074$ & $15.594\pm0.072$ \\
  $\mu_g$ (mag)             & $17.766\pm0.228$ & $15.580\pm0.227$ & $20.109\pm0.231$ & $15.739\pm0.227$ \\
  $\mu_r$ (mag)             & $17.505\pm0.144$ & $15.210\pm0.104$ & $19.909\pm0.146$ & $15.688\pm0.143$ \\
  $\mu_i$ (mag)             & $17.450\pm0.118$ & $15.181\pm0.086$ & $19.862\pm0.121$ & $15.697\pm0.117$ \\
  $\mu_z$ (mag)             & $17.373\pm0.109$ & $\cdots$         & $19.726\pm0.114$ & $\cdots$  \\
  $\mu_y$ (mag)             & $17.449\pm0.124$ & $\cdots$         & $\cdots$         & $\cdots$  \\
  \enddata
  \tablenotetext{a}{$\mathrm{[Fe/H]}_{\mathrm{lit}}$ is adopted from \citet{dekany2022}; $\mu_{\mathrm{lit}}$ is adopted from \citet{stringer2021}.}
  \tablenotetext{b}{Both $\mathrm{[Fe/H]}_{\mathrm{lit}}$ and $\mu_{\mathrm{lit}}$ are adopted from \citet{li2023}.}
  \tablenotetext{c}{$\mathrm{[Fe/H]}_{\mathrm{lit}}$ is adopted from \citet{dekany2022}.}
\end{deluxetable*}

\section{The EDFS, Rubin95 and Rubin38 Field} \label{sec_3field}

We first analyzed the four known RR Lyrae in the EDFS, Rubin95, and Rubin38 field, for which the DP1 light curves and the best-fit template light curves are displayed in Figure \ref{fig_lc1} and \ref{fig_lc38}. Their multiband mean magnitudes and amplitudes returned from the template light curves fitting, as well as other relevant information and results, are summarized in Table \ref{tab_vsx}.

\begin{figure}
  \epsscale{1.1}
  \plotone{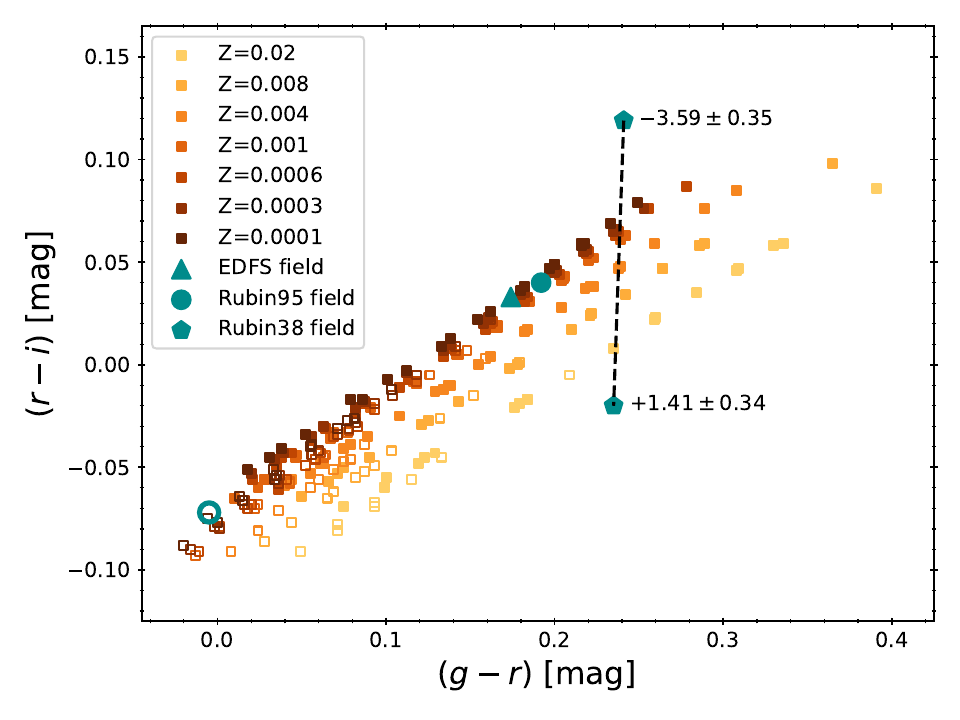}
  \caption{Comparison of the extinction-corrected colors for the four known RR Lyrae listed in Table \ref{tab_vsx} to the metallicity-dependent theoretical color-color tracks (square symbols) based on the pulsation models constructed in \citet{marconi2022}. For the three RR Lyrae in EDFS and Rubin95 field, there is no discernible difference on the colors derived from either the {\tt objectForcedSource} light curves or the {\tt diaObjectForcedSource} light curves, hence we only plotted their colors based the {\tt objectForcedSource} light curves. In case of the RR Lyrae in the Rubin38 field, its $i$-band mean magnitudes are different by $0.143$~mag, resulting in discrepant $(r-i)$~mag colors. Hence this RR Lyrae was plotted twice (and connected by a dashed line). The estimated metallicities were also labeled next to these two points. Filled and open symbols represent the RRab and RRc pulsators (or models), respectively.}
  \label{fig_vsx}
\end{figure}

As can be seen from Table \ref{tab_vsx}, values of the mean magnitudes and amplitudes in various filters are in very good agreements (with a difference of $\sim0.05$~mag or smaller) derived from using the {\tt objectForcedSource} light curves and the {\tt diaObjectForcedSource} light curves. The only exceptions are the $u$-band light curves for GaiaDR3 2912361076526273664 in the Rubin95 field, and the $i$-band light curves for CSS J023424.4+072337 in the Rubin38 field (as discussed in the next subsection). In case of the  RR Lyrae in the EDFS field, the template light curve do not fit well the $r$-band data which has most phase points. Therefore, we compared its light curve to the Gaia \citep{gaia2016} Data Release 3 \citep[DR3,][]{gaia2023} $G$-band light curve data in Figure \ref{fig_edfs}, suggesting this RR Lyrae could be a Blazhko RR Lyrae. 

\subsection{Colors and Metallicity}

These four RR Lyrae have very precise $gri$-band mean magnitudes (derived from fitting the template light curves; see subsection \ref{sec_tmplc}), allowing the photometric metallicities to be estimated using the theoretical $gri$-band colors-metallicity relation presented in \citet[][their Table 5]{marconi2022}, i.e. $\mathrm{[Fe/H]}_{gri}=\alpha (g-r)+\beta (r-i)+\gamma$ (with a dispersion of $0.31$~dex). As quoted from \citet{marconi2022}, such a relation ``appears to be an interesting tool to infer the metal abundance from $g_{LSST}$, $r_{LSST}$, and $i_{LSST}$ photometry with an expected negligible dependence on uncertainties in reddening determinations". In Figure \ref{fig_vsx}, we compared their extinction-corrected colors, derived using the mean magnitudes and $E(B-V)$ values listed in Table \ref{tab_vsx}, to the metallicity-dependent theoretical color-color tracks based on the pulsation models of \citet{marconi2022}. The extinction corrected $(g-r)$ and $(r-i)$ colors for the three RR Lyrae in the EDFS field and the Rubin95 field are located within the low-metallicity tracks defined by the pulsation models (see Figure \ref{fig_vsx}), suggesting these three RR Lyrae have a low metallicity. As can be seen from Table \ref{tab_vsx}, the derived photometric metallicities for the three RR Lyrae in the EDFS and the Rubin95 field have $\mathrm{[Fe/H]}_{gri}\sim -1.8$~dex. Also, they are in good agreement with measurements from the {\tt objectForcedSource} and {\tt diaObjectForcedSource} light curves, due to excellent agreements on their mean magnitudes. 

The colors for the only RR Lyrae in the Rubin38 field (CSS J023424.4+072337), on the other hand, are located outside the theoretical color tracks, mainly caused by the discrepant $i$-band mean magnitudes. These also affect the estimation of $\mathrm{[Fe/H]}_{gri}$ for this RR Lyrae, which give spurious values as listed in Table \ref{tab_vsx}. The left and middle panels of Figure \ref{fig_lc38} compare the best-fit $gri$-band template light curves to the {\tt objectForcedSource} and {\tt diaObjectForcedSource} light curve data for this RR Lyrae, respectively. These panels clearly demonstrate the discrepant best-fit $i$-band template light curves. By fixing the $i$-band amplitude in the template-fit code via the $gi$-band amplitude ratio found in \citet{braga2024}, the best-fit $i$-band template light curve in the right panel is now in very good agreement. However, the derived photometric metallicity is $\mathrm{[Fe/H]}_{gri}\sim +0.90$~dex, which is still unrealistically high for a RR Lyrae. This example demonstrates fitting the template light curves to the sparsely sampled light curves could give discrepant results, and should be treated with caution.   

\subsection{Distance Modulus} \label{sec_mu}

\citet{marconi2022} also derived a set of theoretical PLZ and PWZ relations, as well as the metallicity - absolute $g$-band magnitude ($M_g$-$\mathrm{[Fe/H]}$) relations, in the LSST filters. These relations have a considerable metallicity term varying from $\sim0.1$~mag/dex to $\sim0.4$~mag/dex, with the exceptions of PWZ relations based on the $r$-band mean magnitudes that use either the $(u-r)$ or the $(g-r)$ colors. \citet{marconi2022} did not provide the expression for the former Wesenheit magnitudes, therefore we adopted $W=r-2.796(g-r)$ PWZ relation to calculate the distance modulus ($\mu$) for these four RR Lyrae. The adopted PWZ relation have a metallicity term of $\sim0.05$~mag/dex, such that the impact of metallicity on the calculated $\mu$ is minimal. Using $\mathrm{[Fe/H]}_{gri}$ derived in the previous section, we calculated $\mu_W$ that are provided in Table \ref{tab_vsx} for the four RR Lyrae. Again, apart from the RR Lyrae in the Rubin38 field (due to spurious $\mathrm{[Fe/H]}_{gri}$), $\mu_W$ calculated using either the {\tt objectForcedSource} light curves or the {\tt diaObjectForcedSource} light curves are in excellent agreement.     

Similarly, we calculated $\mu_\lambda$ in $grizy$ bands using the theoretical PLZ and $M_g$--$\mathrm{[Fe/H]}$ relations (with the same $\mathrm{[Fe/H]}_{gri}$). \citet{marconi2022} provided a linear and a quadratic version of the $M_g$--$\mathrm{[Fe/H]}$ relation. We adopted the linear version because the $\mu_g$ calculated using both versions are consistent with each other. The calculated $\mu_\lambda$ using the extinction-corrected mean magnitudes are summarized in Table \ref{tab_vsx}. We noticed several trends in these $\mu_\lambda$ -- firstly, all $\mu_\lambda$ are larger than $\mu_W$, and these estimates at longer wavelengths are in better agreement with $\mu_W$ than at the short wavelengths. Furthermore, the values of $\mu_\lambda$ decrease as $\lambda$ increases. This trend is unlikely to be caused by extinction, because the same trend is seen for the RR Lyrae in the lowest extinction field (the EDFS field) as well. We discuss a possible reason causing these trends in the last section.

\subsection{Compare to Literature Values}

In Table \ref{tab_vsx}, we have also collected literature photometric metallicity ($\mathrm{[Fe/H]}_{\mathrm{lit}}$) and distance modulus ($\mu_{\mathrm{lit}}$), whenever available, measured from recent works for the four RR Lyrae. These estimates serve as reference for a relative comparison with the values measured from the DP1 light curves. We emphasize that these literature values should not be treated as the ``true values'', as they might also suffer from various systematic errors. Nevertheless, these literature values are completely independent of DP1. 

Among the four RR Lyrae, both $\mu_W$ and $\mathrm{[Fe/H]}_{gri}$ for the only RR Lyrae in the EDFS field are in agreement with the literature values. For the RRab star in the Rubin95 field, its measured $\mathrm{[Fe/H]}_{gri}$ is still consistent within $\sim 1.5\sigma$ with the literature value. The values of $\mathrm{[Fe/H]}_{\mathrm{lit}}\sim -2.2$~dex for these two RRab stars, both measured based on a deep learning approach \citep{dekany2022}, are more metal-poor than $\mathrm{[Fe/H]}_{gri}\sim -1.8$~dex, implying a possible systematic offset between the two methods.   

In contrast, there is a $\sim 1$~dex or $\sim 2.5\sigma$ difference between $\mathrm{[Fe/H]}_{\mathrm{lit}}$ and $\mathrm{[Fe/H]}_{gri}$ for the RRc star in the Rubin95 field. This could be due to the sparse $i$-band DP1 light curve for this RR Lyrae, as they have less data points than the $gr$-band (see Table \ref{tab_vsx} and the middle of Figure \ref{fig_lc1}). If we adopt $\mathrm{[Fe/H]}_{\mathrm{lit}}$ for this RR Lyrae and recalculate $\mu_W$, we obtain $\mu_W\sim 15.07\pm0.06$~mag, in agreement to the literature value. Similarly, by adopting $\mathrm{[Fe/H]}_{\mathrm{lit}}$ for the RR Lyrae in the Rubin38 field, we obtain $\mu_W\sim 15.76\pm0.08$~mag, or $\sim 1.6\sigma$ larger than the literature estimate.

\begin{figure*}
  \epsscale{1.1}
  \centering
  \begin{tabular}{ccc}
    \includegraphics[width=5.75cm]{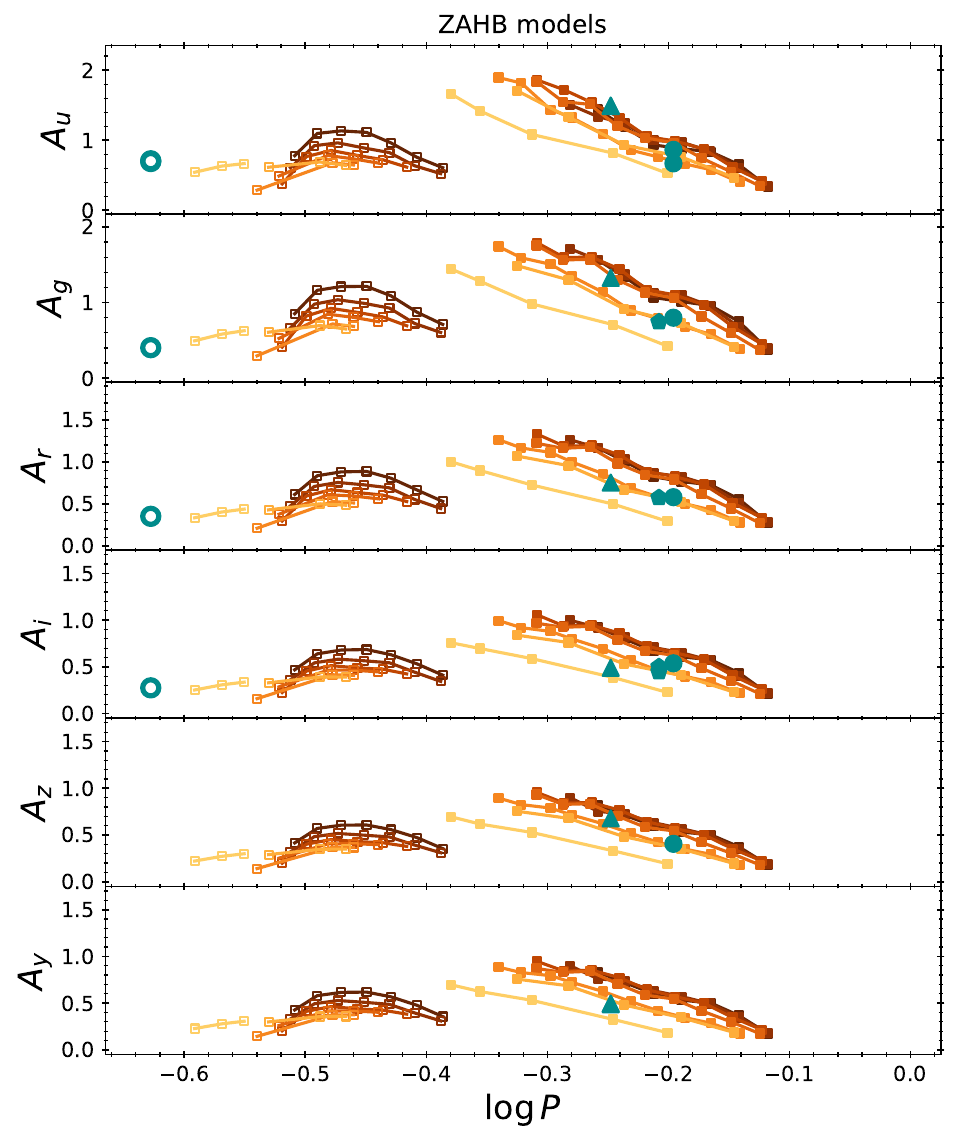} & \includegraphics[width=5.75cm]{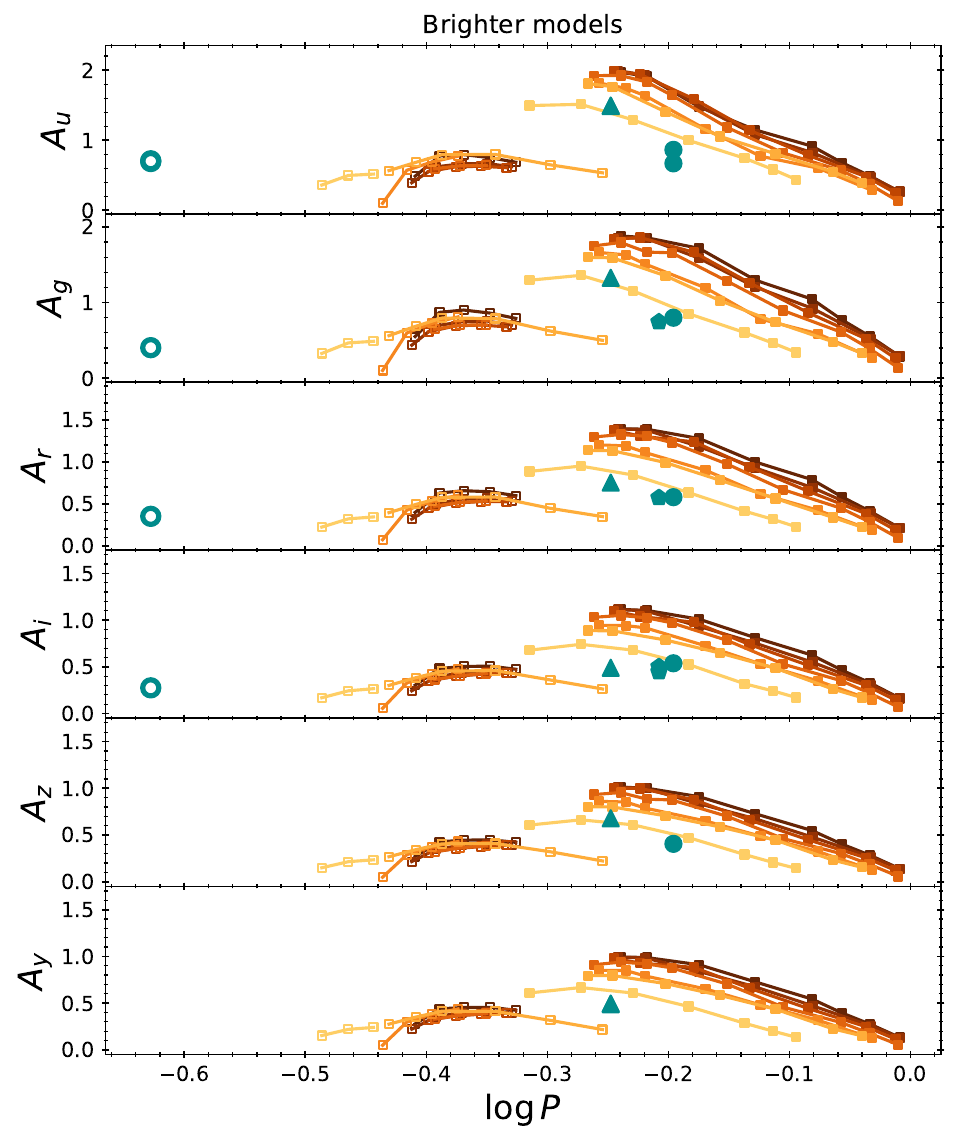} & \includegraphics[width=5.75cm]{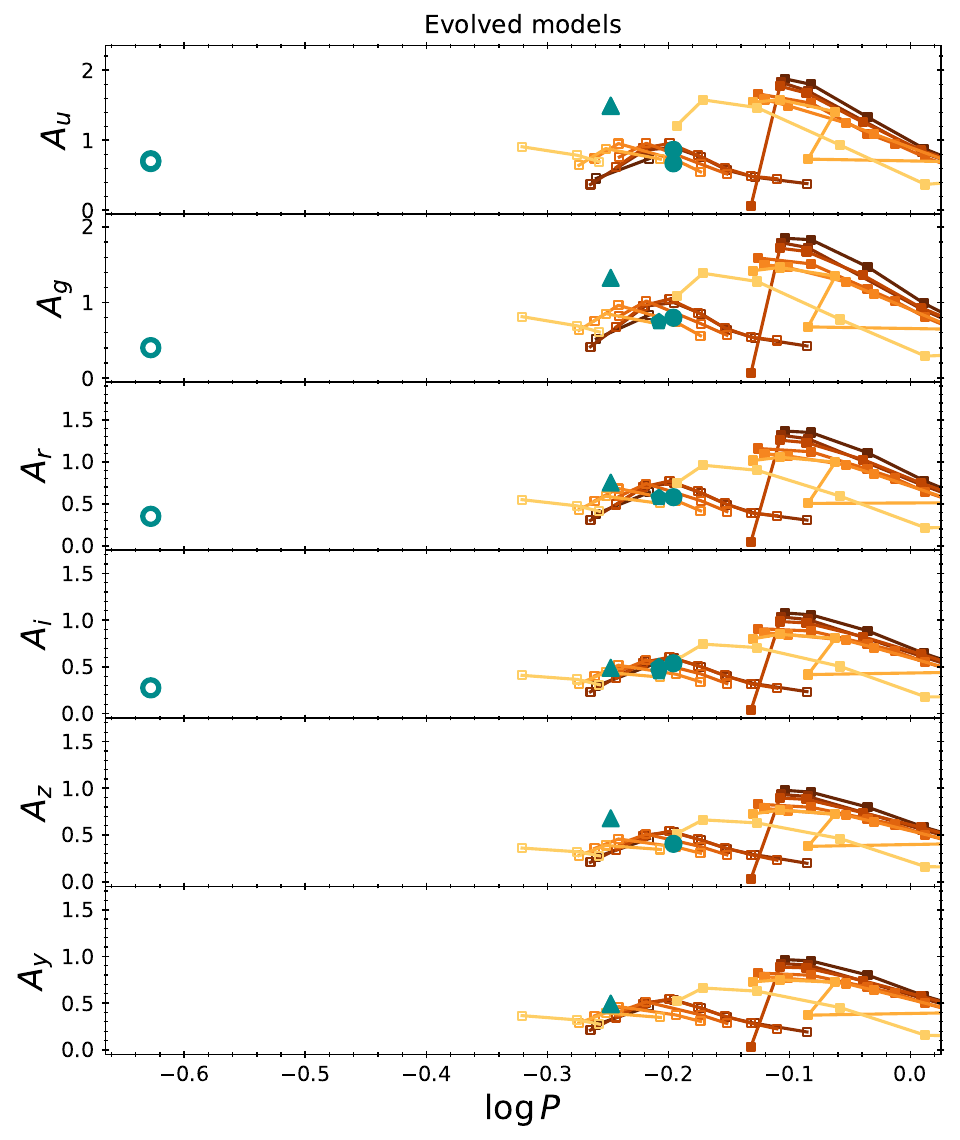} \\
  \end{tabular}
  \caption{Comparison of the light curve amplitudes listed in Table \ref{tab_vsx} to the theoretical Bailey diagrams, separately for the three sets of input pulsation models \citep[for more details, see][]{marconi2022}. Symbols and the color codes are same as in Figure \ref{fig_vsx}.}
  \label{fig_ampvsx}
\end{figure*}

\subsection{Amplitudes}

In \citet{marconi2022}, the authors considered three sets of RR Lyrae pulsating models: the zero-age horizontal branch (ZAHB) models, the brighter models, and the evolved models. In the period-amplitude diagram, or the Bailey diagram, all of these three sets of models predict smaller amplitudes (at a fixed period) for higher metallicity models. However, the predicted period ranges were shifted to a longer period for both the brighter and the evolved models \citep[see Figure 3 \& 4 in][]{marconi2022}.

In Figure \ref{fig_ampvsx}, we compared the multiband amplitudes for these four RR Lyrae to the theoretical Bailey diagrams. In general, the observed amplitudes of the three RRab stars follow similar trend of decreasing amplitudes with increased periods (except the $i$-band amplitudes for the RRab star in the EDFS field). The amplitudes of the RRc star in the Rubin95 field are also consistent with the values of the theoretical amplitudes, albeit it has a shorter period.      

As can be seen from Figure \ref{fig_ampvsx}, the observed amplitudes for the RRab stars are consistent with the theoretical amplitudes based on the ZAHB models, and totally inconsistent with the evolved models. In case of the brighter models, the observed amplitudes are located along, or close to, the theoretical track with $Z=0.02$, which is unlikely given that these RR Lyrae are metal-poor stars. 

\begin{figure*}
  \epsscale{1.1}
  \plottwo{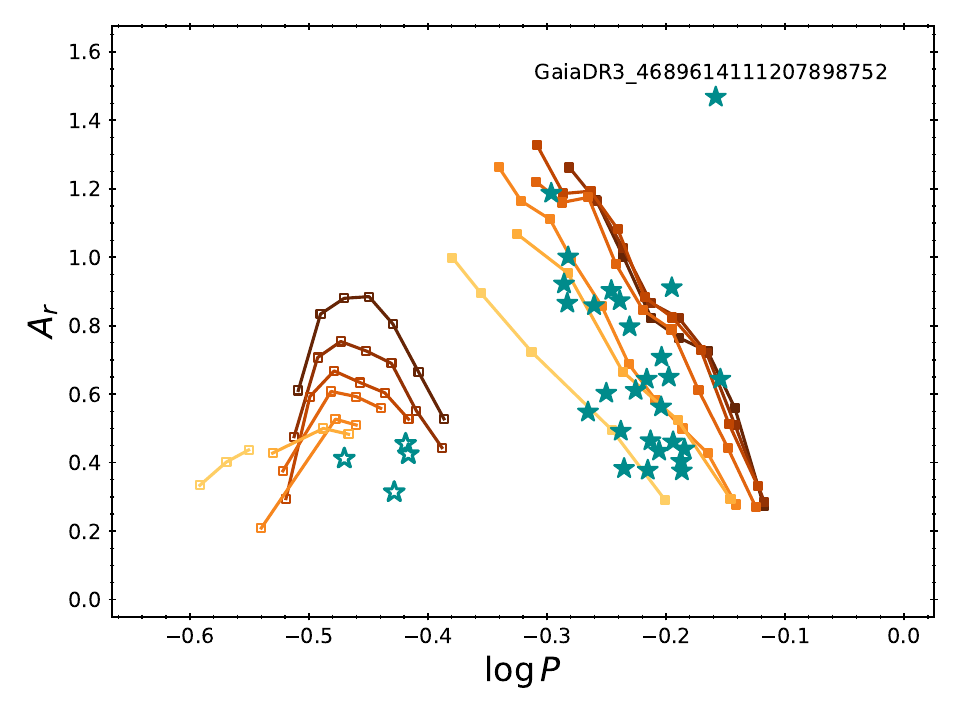}{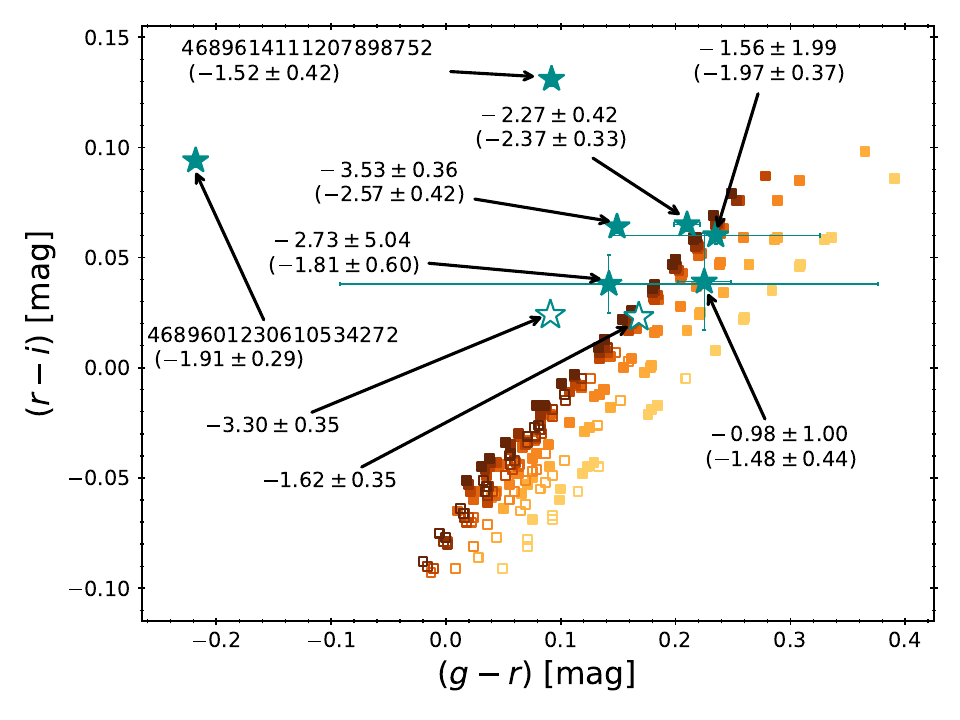}
  \caption{{\it Left Panel:} Comparison of the $r$-band amplitudes for the known RR Lyrae in the 47Tuc field (star symbols) to the theoretical Bailey diagrams based on the ZAHB models \citep{marconi2022}. The $r$-band light curve for the outlier marked in the plot can be found in the upper-left panels of Figure \ref{fig_47tuc_lc}. {\it Right Panel:} Same as Figure \ref{fig_vsx}, but for the nine RR Lyrae with $gri$-band light curves. The estimated $\mathrm{[Fe/H]}_{gri}$, using the theoretical color-color relation presented in \citet{marconi2022}, are annotated for each RR Lyrae except the two extreme outliers. These two extreme outliers have unrealistic $\mathrm{[Fe/H]}_{gri} < -4$~dex, hence we only provided their Gaia DR3 ID numbers. Values in the parentheses are the literature $\mathrm{[Fe/H]}$ adopted from \citet{li2023}. In both panels, color codes for the theoretical models are same as in Figure \ref{fig_vsx}. Filled and open symbols represent the RRab and RRc stars (or models), respectively.}
  \label{fig_47tuc}
\end{figure*}

\section{The 47Tuc and Fornax Field} \label{sec_2field}

Even though the 47Tuc field and the Fornax fields have more known RR Lyrae than other three DP1 fields, these two fields suffer problems of crowding and smaller numbers of observed epochs (see Table \ref{tab1}). As a result, the DP1 light curves for the RR Lyrae located in these two fields are very sparse and should be treated with caution.

When fitting the template light curves to the RR Lyrae in these two fields, we have also fixed the amplitudes of the $gi$-band light curves (whenever available) with respect to the $r$-band  by using the amplitude ratios provided in \citet{braga2024}. This is because majority of the RR Lyrae in these two fields have the $r$-band light curves, and in general, the $r$-band light curves have the most data-points when compared to the $gi$-band. If only a single band light curve is available for a given RR Lyrae, we did not fix the amplitudes. While fitting the template light curves, we have also excluded those light curves that returned a mean magnitude error larger than 0.5~mag. The final numbers of RR Lyrae are listed in the last column of Table \ref{tab1}. 

\subsection{47Tuc Field: Colors, Metallicity and Amplitudes}

There is only one RR Lyrae left in the 47Tuc field which has both the {\tt objectForcedSource} light curves and the {\tt diaObjectForcedSource} light curves, but only in the $i$-band. Similar to the RR Lyrae in the Rubin38 field, the fitted $i$-band template light curves are discrepant (with a difference of 0.12~mag in the mean magnitudes), hence we did not consider this RR Lyrae further in the analysis.

In contrast, there are 34 RR Lyrae in the 47Tuc field with {\tt diaObjectForcedSource} light curves. In the left panel of Figure \ref{fig_47tuc}, we compared the $r$-band amplitudes for 31 of them\footnote{The rest of the RR Lyrae do not have the $r$-band light curves.} with the theoretical Bailey diagram. Similar to the RR Lyrae in the EDFS, Rubin95, and Rubin38 fields, distribution of the $r$-band amplitudes for these RR Lyrae follows the theoretical trends, except for an outlier RRab star and RRc stars (they seems to have smaller amplitudes). This is likely due to the sparse nature of the DP1 light curves, and will be improved in the future LSST data releases.       

\begin{figure*}
  \centering
  \begin{tabular}{ccc}
    \includegraphics[width=5.75cm]{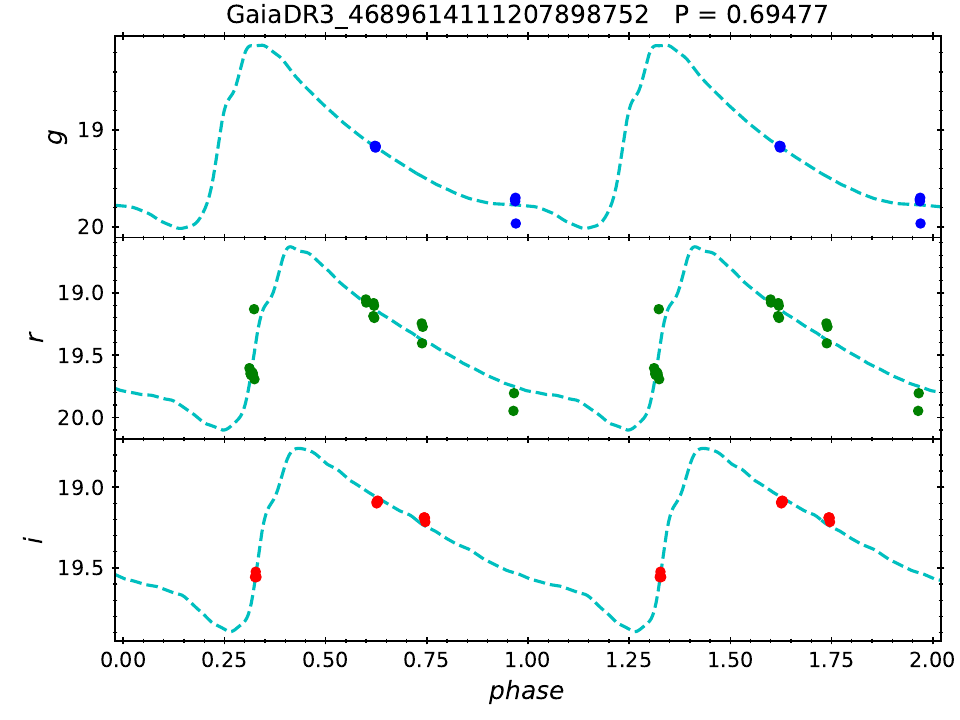} & \includegraphics[width=5.75cm]{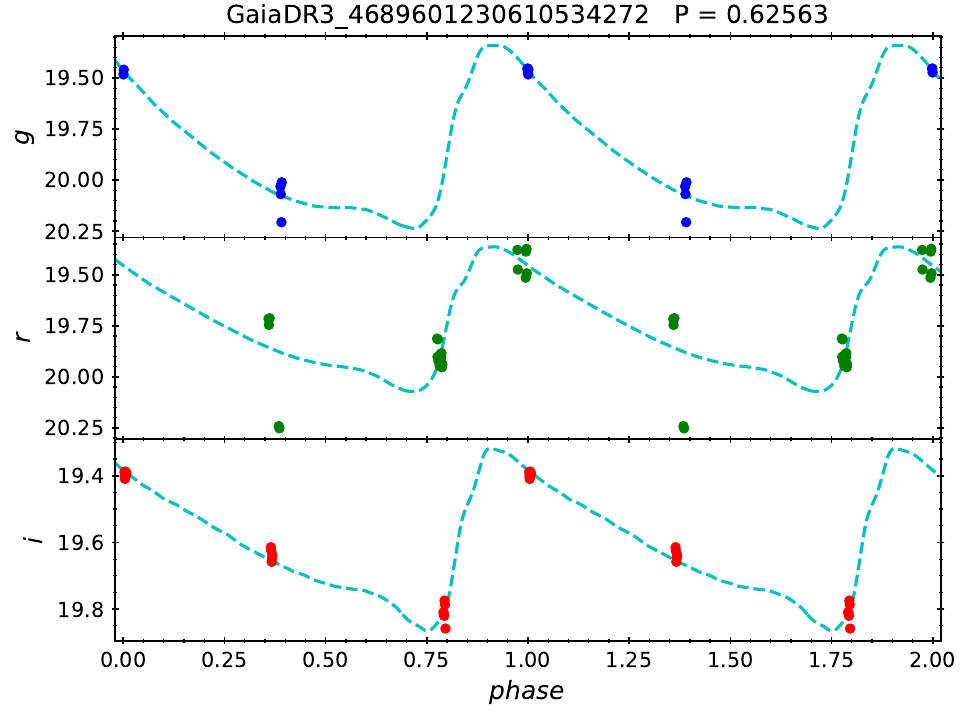} & \includegraphics[width=5.75cm]{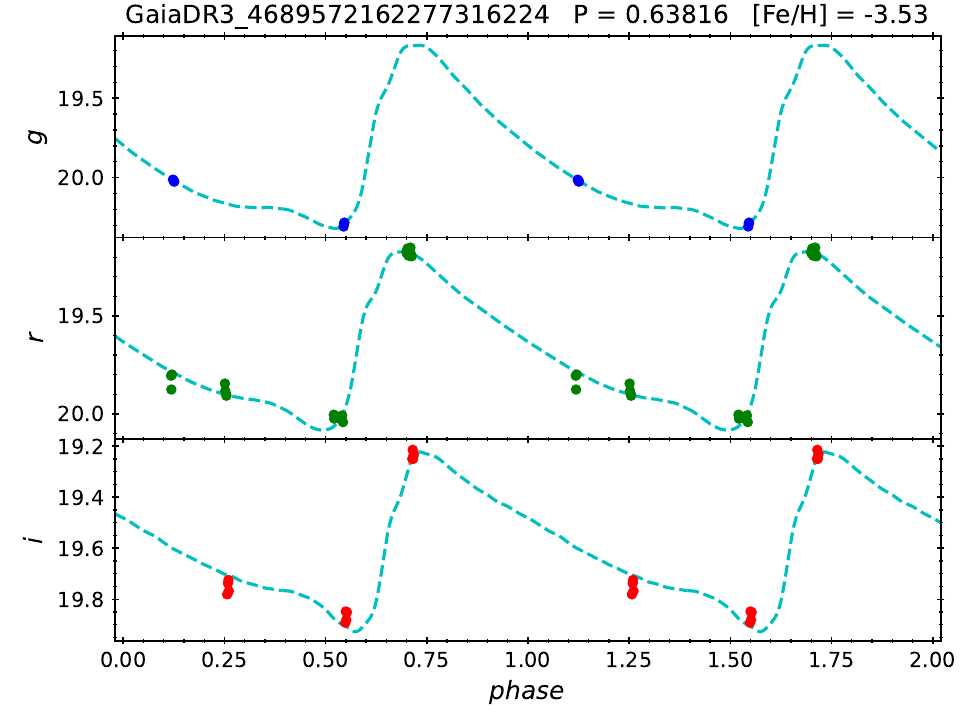} \\
    \includegraphics[width=5.75cm]{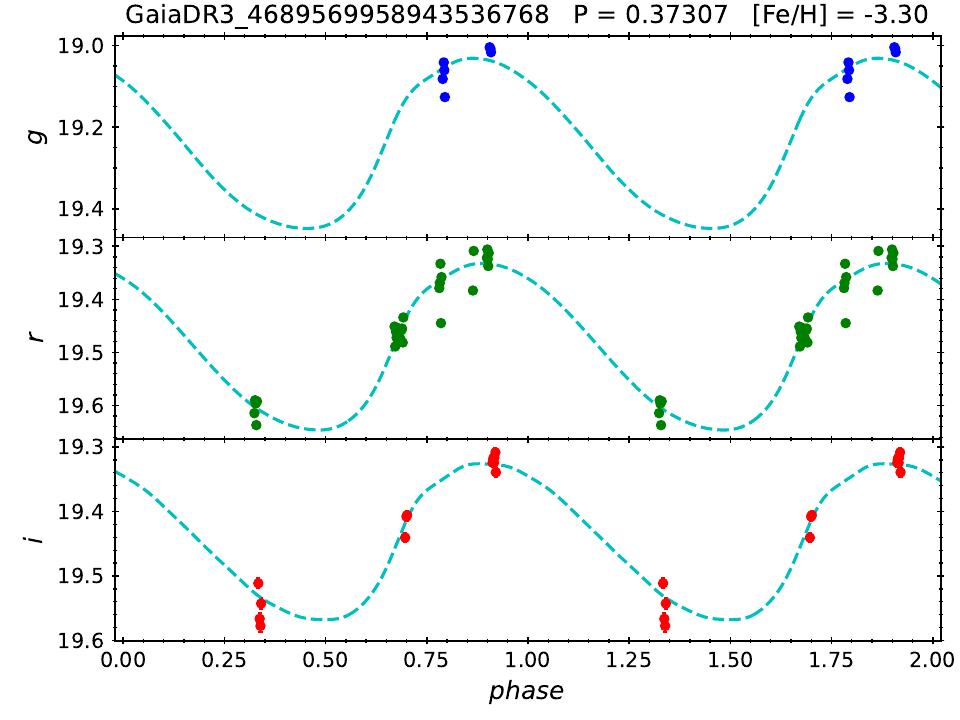} & \includegraphics[width=5.75cm]{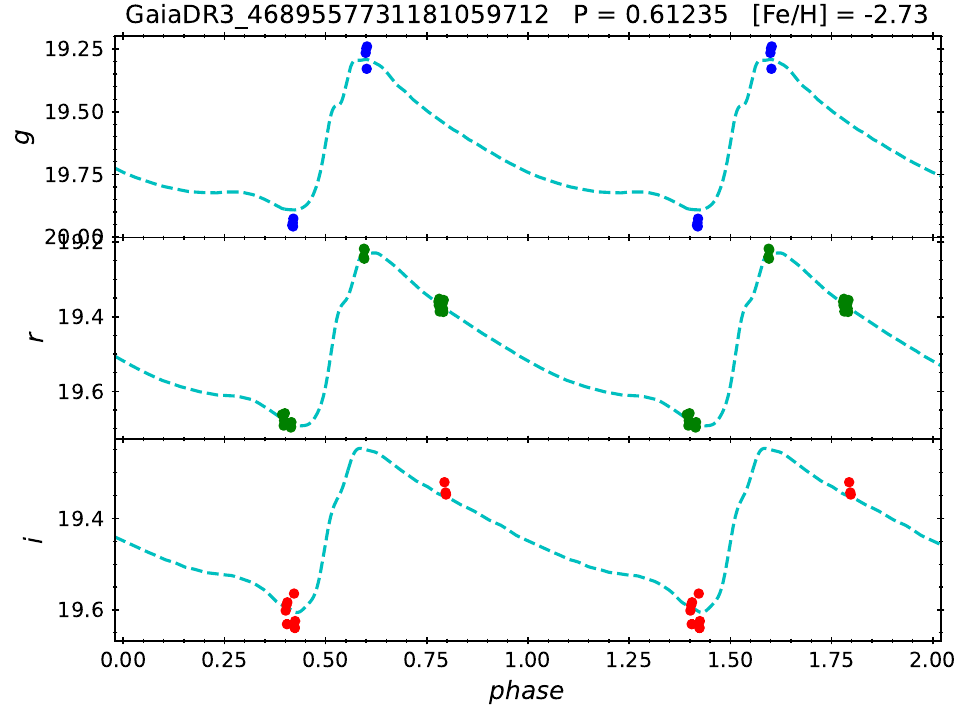} & \includegraphics[width=5.75cm]{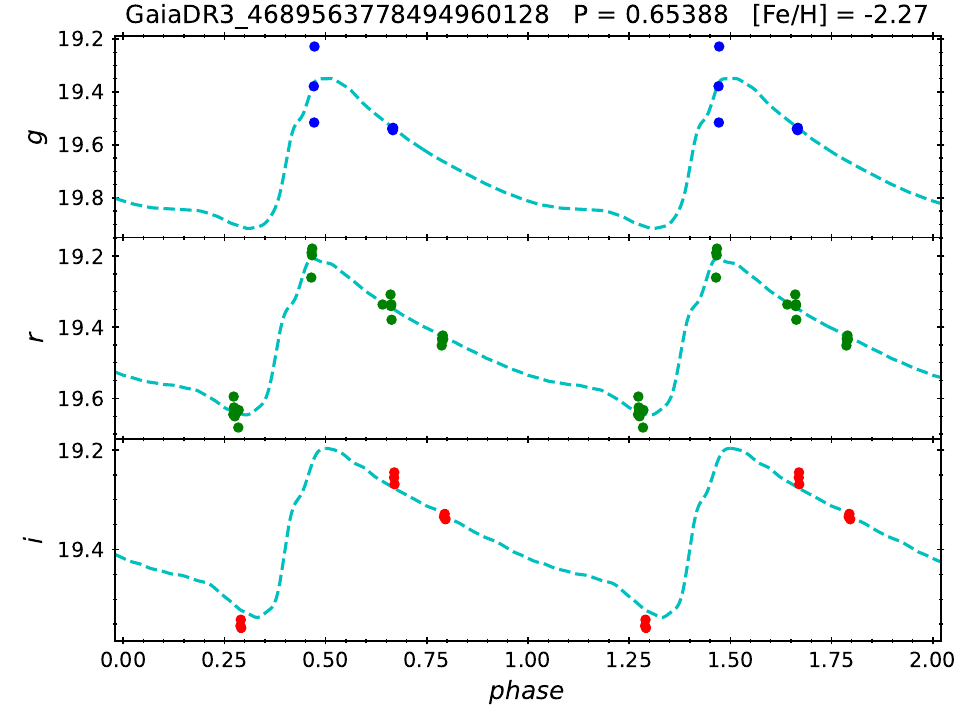} \\
    \includegraphics[width=5.75cm]{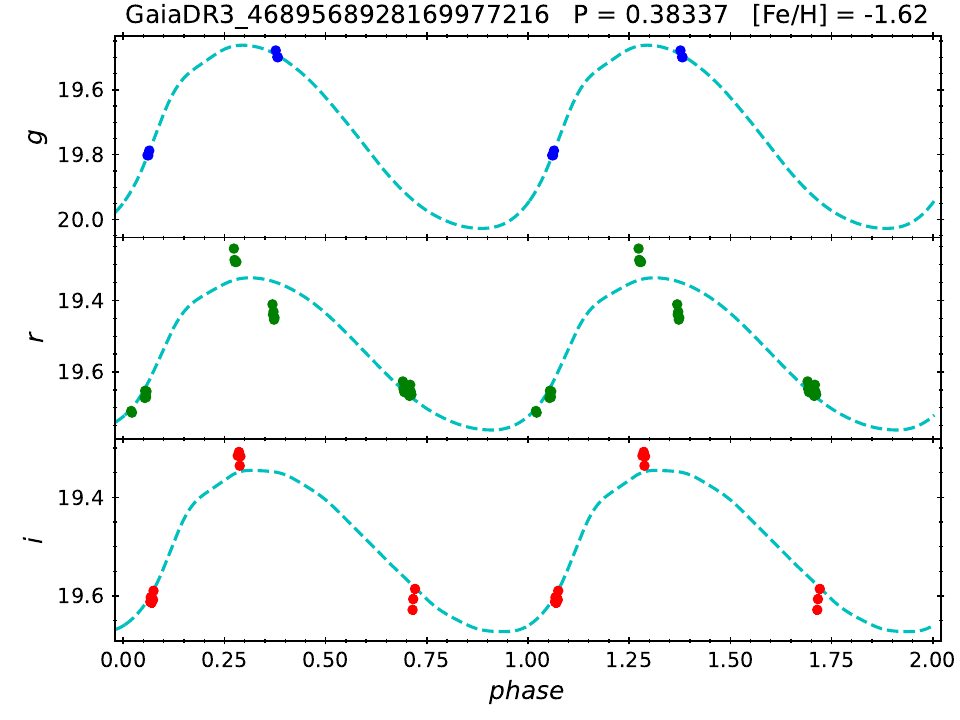} & \includegraphics[width=5.75cm]{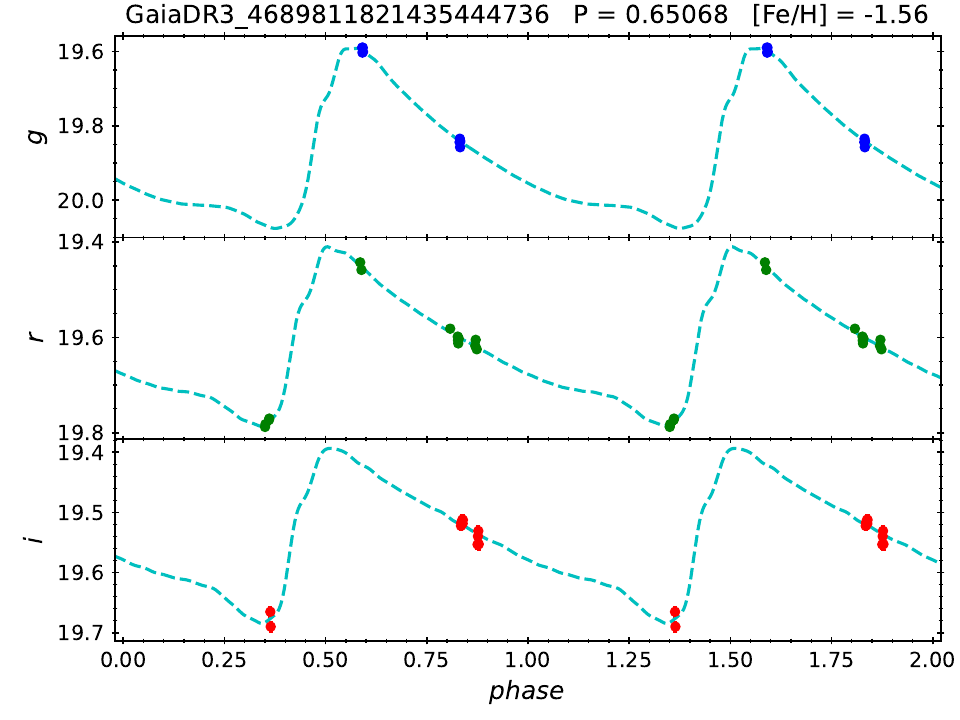} & \includegraphics[width=5.75cm]{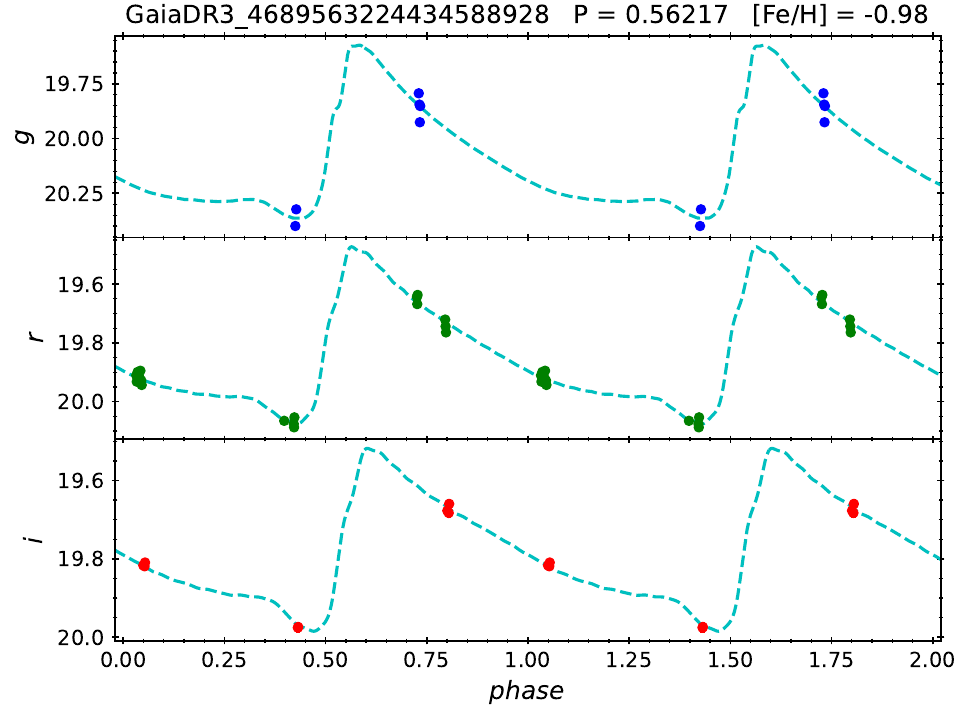} \\
  \end{tabular}
  \caption{The $gri$-band {\tt diaObjectForcedSource} light curves for the nine RR Lyrae in the 47Tuc field. The cyan curves represent the best-fit template light curves. The units for period $P$ and metallicity $\mathrm{[Fe/H]}$ are in days and dex, respectively.}
  \label{fig_47tuc_lc}
\end{figure*}

The sparse DP1 light curves also have an impact on metallicity estimation. Among these RR Lyrae in the 47Tuc field, nine have $gri$-band light curves (as shown in Figure \ref{fig_47tuc_lc}), allowing the estimation of $\mathrm{[Fe/H]}_{gri}$ based on their extinction-corrected colors. The right panel of Figure \ref{fig_47tuc} presents the locations of these nine RR Lyrae in the color-color plane, as well as their estimated $\mathrm{[Fe/H]}_{gri}$ values. A few of them have $\mathrm{[Fe/H]}_{gri}$ consistent with published values from \citet{li2023}. 

\begin{figure*}
  \epsscale{1.1}
  \centering
  \begin{tabular}{ccc}
    \includegraphics[width=5.75cm]{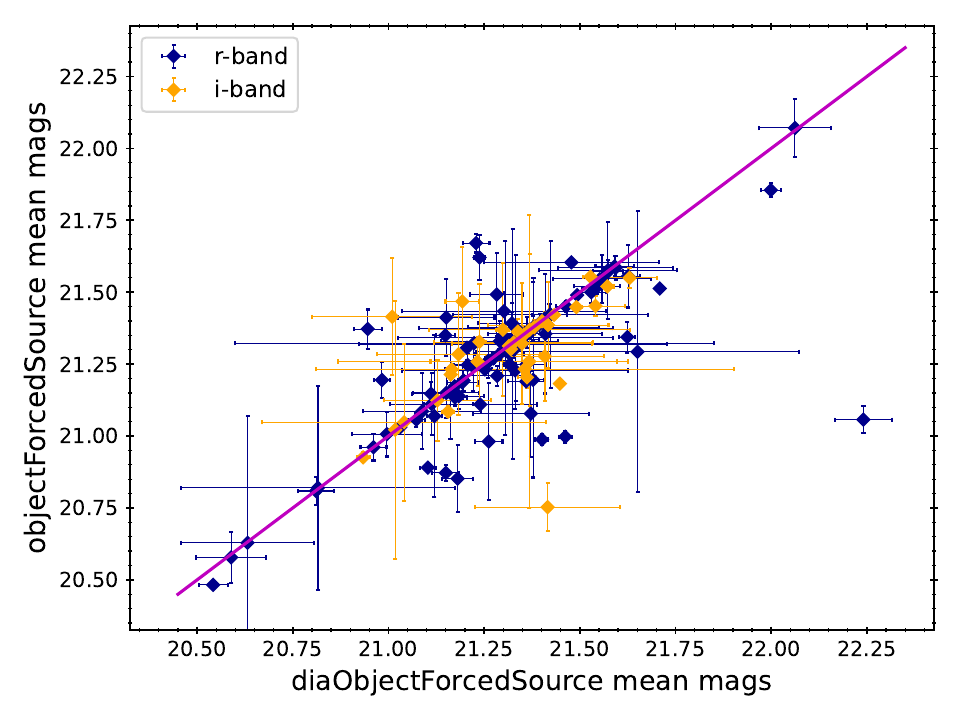} & \includegraphics[width=5.75cm]{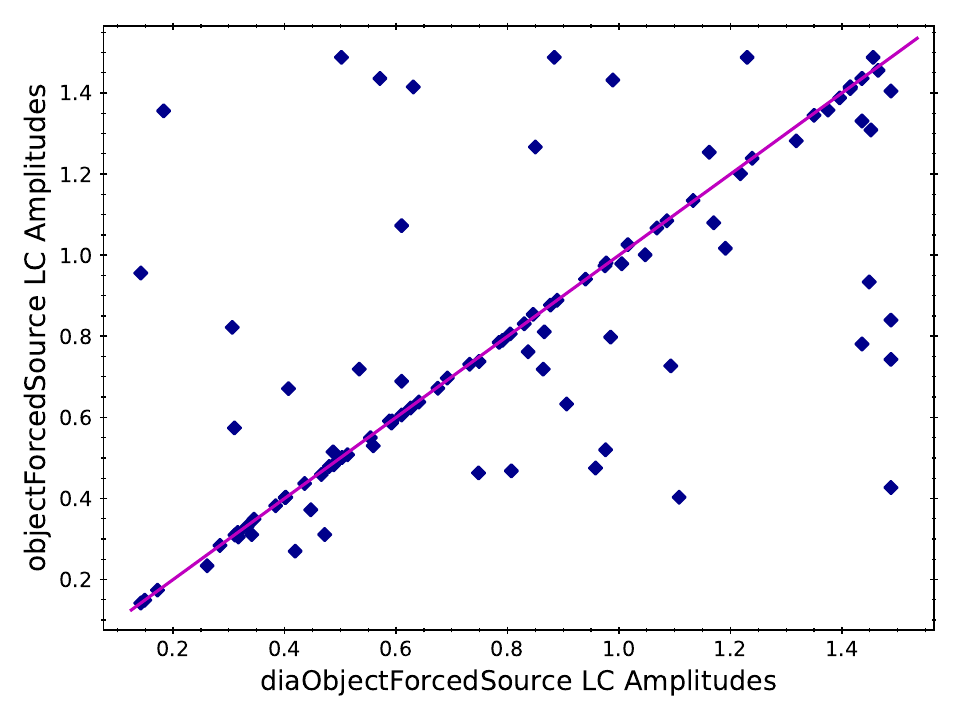} & \includegraphics[width=5.75cm]{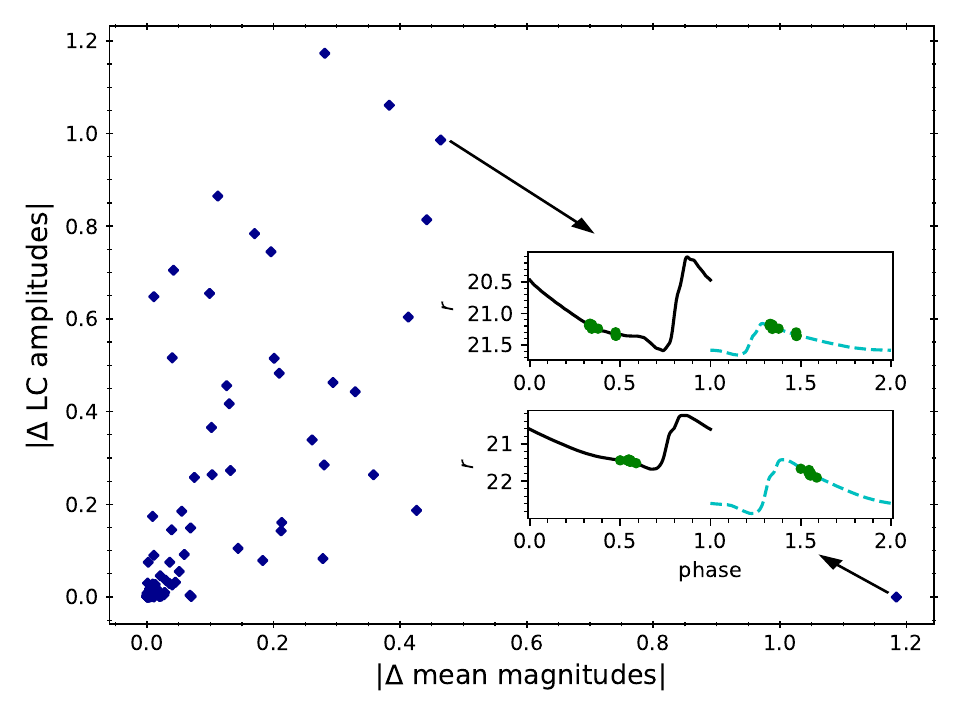} \\
  \end{tabular}
  \caption{{\it Left Panel:} Comparison of the mean magnitudes returned from fitting the {\tt objectForcedSource} light curves and the {\tt diaObjectForcedSource} light curves with the template light curves to the known RR Lyrae in the Fornax field. The solid magenta line represents the 1:1 relation. {\it Middle Panel:} Similar to the left panel, but for the $r$-band amplitudes determined from the best-fit template light curves. We did not compare the $i$-band amplitudes because they are fixed using the $ri$-band amplitude ratio found in \citet{braga2024}. {\it Right Panel:} Correlations between the absolute difference ($|\Delta|$) between the $r$-band mean magnitudes and amplitudes. Light curves for two outlier points are shown in the inset figures, with black solid (from phase 0 to 1) and cyan dashed (from phase 1 to 2) curves represent the best-fit template light curves to the {\tt objectForcedSource} light curves data and the {\tt diaObjectForcedSource} light curves data, respectively.}
  \label{fig_fornaxma}
\end{figure*}

\subsection{Fornax Field: Mean Magnitudes and Amplitudes}

Even though the Fornax field has the least number of epochs, it contained the most number of known RR Lyrae within the seven DP1 fields. Furthermore, about 108 of these RR Lyrae have both the {\tt objectForcedSource} light curves and the {\tt diaObjectForcedSource} light curves, allowing a straight forward comparison of the mean magnitudes and amplitudes based on these two sets of light curves (in addition to those listed in Table \ref{tab_vsx}). The left panel of Figure \ref{fig_fornaxma} compares the $ri$-band mean magnitudes for these two set of light curves, returned from the template light curve fittings, with averaged differences of $\sim 0.01$~mag and $\sim 0.03$~mag in the $ri$-band, respectively. The corresponding dispersion in both filters is $\sim0.10$~mag. Similarly, the average difference of the $r$-band amplitudes is $0.00$~mag, with a dispersion of $0.32$~mag. Comparison of the $r$-band amplitudes is shown in the middle panel of Figure \ref{fig_fornaxma}. Finally, the right panel of Figure \ref{fig_fornaxma} presents the correlation between the absolute difference of mean magnitudes and amplitudes. As expected, light curves with larger difference in mean magnitudes also tend to have larger difference in amplitudes, due to the nature of sparsely sampled light curves (with examples shown in the inset figures of Figure \ref{fig_fornaxma}).

\subsection{Distance Modulus Comparison}

\begin{figure*}
  \epsscale{1.1}
  \plottwo{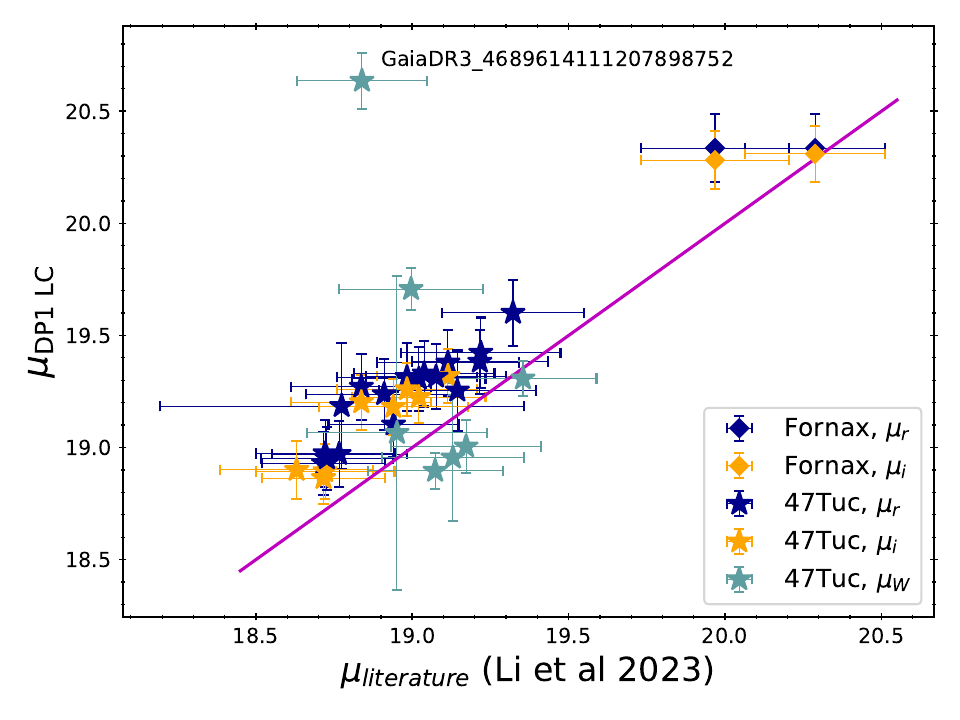}{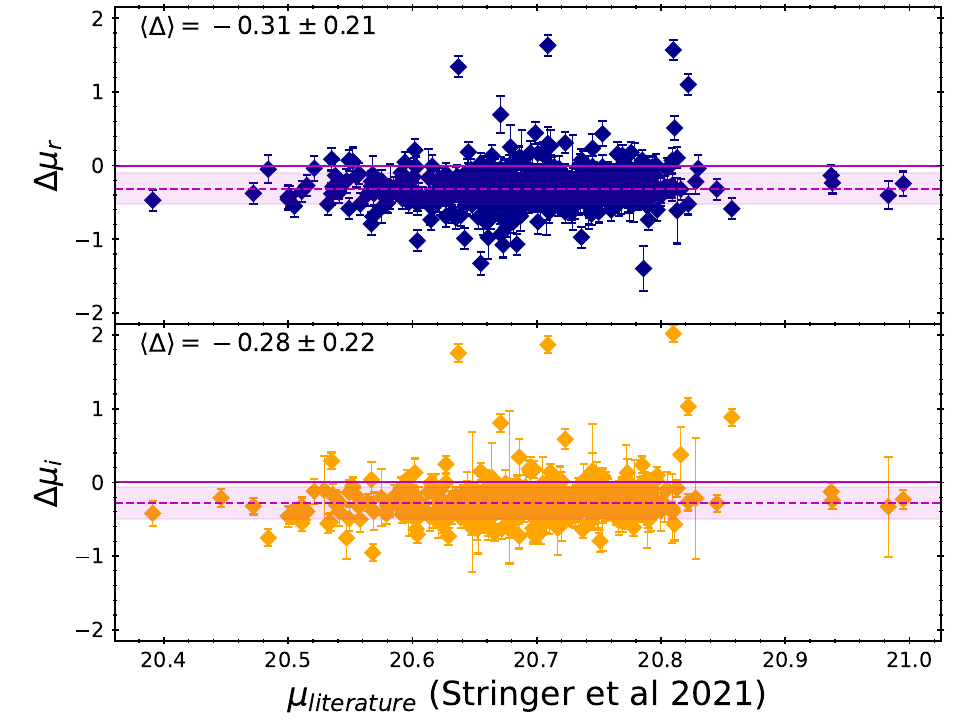}
  \caption{{\it Left Panel:} Comparison of the derived $\mu$, using the mean magnitudes obtained from the {\tt diaObjectForcedSource} light curves and the theoretical PWZ and PLZ relations \citep{marconi2022}, to the literature values adopted from \citet[][whenever available]{li2023}. The straight line represents the $1:1$ relation. {\it Right Panel:} Similar to the left panel, but for the RR Lyrae in the DES sample with literature values adopted from \citet{stringer2021}, where $\Delta = \mu_{\mathrm{literature}} - \mu_{ri}$. The solid lines are for $\Delta =0$. The dashed lines represent the averaged $\Delta$ (after excluding outliers with $|\Delta| > 1.0$~mag) and shaded areas are $\pm1\sigma$ dispersion of the averaged values (given on the top-left of each panels). Results from the {\tt objectForcedSource} light curves are similar and are not shown.}
  \label{fig_mu}
\end{figure*}

Similar to subsection \ref{sec_mu}, we derived $\mu$ for RR Lyrae in both of the 47Tuc and the Fornax field, using the theoretical PLZ or PWZ relations presented in \citet{marconi2022}, and compare them to the literature values. In case of the RR Lyrae in the VSX sample, we applied the same PWZ relation as in subsection \ref{sec_mu} to the nine 47Tuc RR Lyrae which have $gri$-band light curve data (i.e., those presented in Figure \ref{fig_47tuc_lc}). For other RR Lyrae in the VSX and the DES samples, we derived their $\mu_{ri}$ (whenever applicable) separately using the PLZ relations after correcting their $ri$-band mean magnitudes with foreground extinction. The $\mathrm{[Fe/H]}_{gri}$ estimates may not be reliable (or not even available) for the RR Lyrae in these two samples. Therefore, we adopted the $\mathrm{[Fe/H]}$ values from \citet[][whenever available]{li2023} and set $\mathrm{[Fe/H]}=-2.10\pm0.27$~dex \citep{braga2022} for the RR Lyrae in the VSX and the DES samples, respectively. Similarly, we adopted the $\mu$ values from \citet{li2023} and \citet{stringer2021} for the VSX and the DES samples, respectively, for a comparison with the DP1-based $\mu$ values.    

Comparisons of $\mu$ valuesfor those RR Lyrae in the VSX sample with literature values are presented in the left panel of Figure \ref{fig_mu}. On average, the derived $\mu_W$ based on the DP1 light curves and \citet{marconi2022} PWZ relation are consistent with the literature values with an offset of $0.05$~mag. This is after excluding the obvious outlier marked in the left panel of Figure \ref{fig_mu} (its light curves can be found in Figure \ref{fig_47tuc_lc}). In contrast, the $\mu_{ri}$ values are larger than the literature values by $\sim 0.25$~mag and $\sim 0.22$~mag in the $ri$ bands, respectively. A similar result was also found in the DES sample, in a sense that $\mu_{ri}$ based on the DP1 light curves and theoretical PLZ relations tend to be larger than the literature values (see the right panel of Figure \ref{fig_mu}). We recall that this trend is also seen for the RR Lyrae listed in Table \ref{tab_vsx}.

\section{Discussion and Conclusion} \label{sec_last}

In this work, we investigated the known RR Lyrae in five (out of seven) DP1 fields by deriving their mean magnitudes and amplitudes (via a template light curves fitting approach) from the DP1 light curves. These RR Lyrae can be separated into two groups. The first group consists of three RR Lyrae with well-sampled light curves, with more than five epoch of observations (at least in the $gri$-band), which are located in the EDFS and the Rubin95 fields. Another group of RR Lyrae includes those located in the Rubin38, the 47Tuc, and the Fornax fields, with (very) sparse DP1 light curves from five or less epoch of observations, and only available in the $gri$-band or even less number of filters. Certainly, as emphasized in th \citet{RTN-095}, the duration, cadence, and number of observed epochs on these DP1 fields do not resemble the official and final LSST observing strategy. Nevertheless, these two groups provide insights to the RR Lyrae light curve quality in early LSST operations (when only a few epochs of data is available) and after accumulating a few years of observations with well sampled light curves.     

Using the LSDB framework, the DP1 data consist of two sets of light curves, the {\tt objectForcedSource} light curves and the {\tt diaObjectForcedSource} light curves, on the detected variable (and transient) sources. \citet{RTN-095} recommended using the {\tt diaObjectForcedSource} light curves for time-domain studies (including variable stars), because fluxes in the {\tt objectForcedSource} light curves could be contaminated by fluxes emitted from nearby sources. In addition, the {\tt DiaObject} catalog could recover more variable sources than the {\tt Object} catalog, at least for the RR Lyrae as demonstrated in \citet{choi2025} and in Table \ref{tab1}. Nevertheless, we recommend to include the {\tt objectForcedSource} light curves for RR Lyrae-related studies. For well-sampled light curves, there is no discernible difference on the mean magnitudes and amplitudes obtained from these two sets of light curves. On the other hand, comparing their mean magnitudes (and amplitudes) from sparsely observed light curves could be used to diagnose or capture ``unreliable'' light curves (such as the example showed for the $i$-band light curve for the RR Lyrae in the Rubin38 field).  

We have also compared or applied various theoretical results derived in \citet{marconi2022} to the extinction-corrected mean magnitudes and hence colors. In particular, the extinction-corrected colors in the $gri$-band can then be used to estimate $\mathrm{[Fe/H]}_{gri}$ using the theoretical color-color relation. Clearly, poor or sparsely sampled light curves, even only just in one band, could produce spurious or unreliable value for the metallicity, as demonstrated for (some of) the RR Lyrae in the Rubin38 and the 47Tuc field. In contrast, the $\mathrm{[Fe/H]}_{gri}$ derived for the two RRab stars in the EDFS and the Rubin95 field, with well-sampled $gri$-band light curves, are consistent with the literature values. Our tests show that the estimation of photometric metallicity from $gri$-band colors is sensitive to the derived mean magnitudes (which needs to be precise and accurate) in all three bands and is possibly prone to additional systematic biases.

As noted in \citet{marconi2022}, one characteristic or advantage of using the color-color relation to estimate $\mathrm{[Fe/H]}_{gri}$ is that the slope of the relation is nearly parallel to the reddening vector defined by the popular extinction law of \citet{cardelli1989}. However, recent studies \citep[e.g.,][]{gordon2023} showed that the reddening law could vary along various sight lines in the Milky Way, which could add an additional systematic error to the estimated $\mathrm{[Fe/H]}_{gri}$ for those RR Lyrae located in the Galactic plane. Note that the DP1 fields, except the Seagull field, are located away from the Galactic plane, and hence the impact of varying reddening law is somewhat minimized in the analysis presented in this work.

\begin{figure}
  \epsscale{1.1}
  \plotone{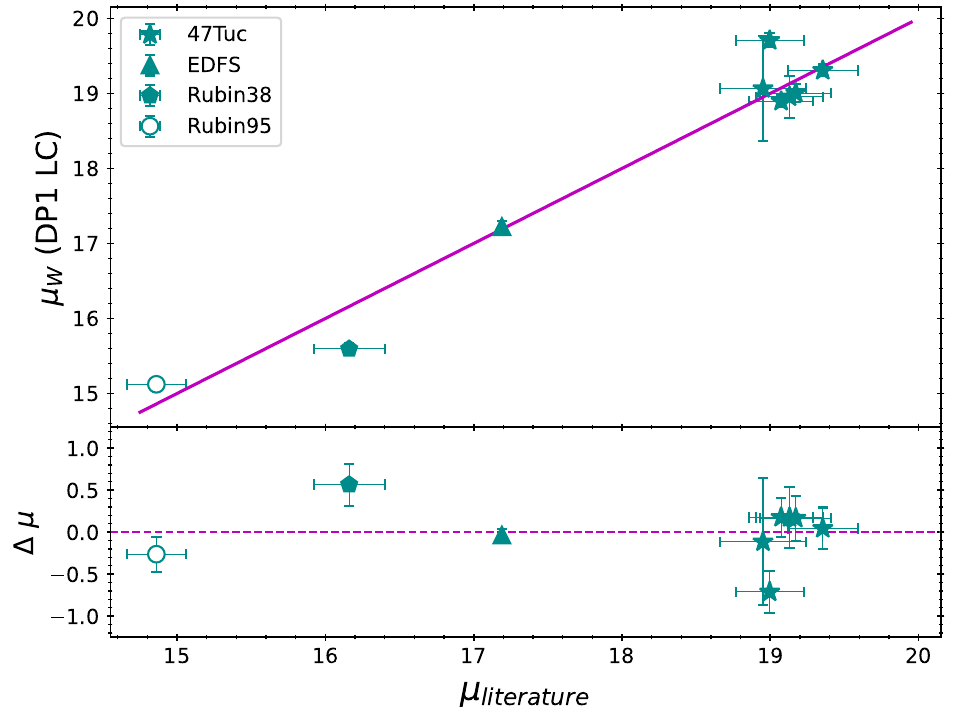}
  \caption{Comparison of $\mu_W$ derived in this work with literature values, including the eight RR Lyrae in 47Tuc field (after excluding the extreme outlier as shown in the left panel of Figure \ref{fig_mu}), and the RR Lyrae listed in Table \ref{tab_vsx}. For consistency, we only show the $\mu_W$ derived from the {\tt diaObjectForcedSource} light curves. The solid line and dashed line in the top and bottom panel represents the $1:1$ relation and $\Delta \mu = 0$, respectively. The weighted mean of $\Delta \mu$ is $-0.01\pm0.36$~mag.}
  \label{fig_muw}
\end{figure}

In case of distance moduli derived from the theoretical PWZ and PLZ relations, we note that, in general, they are larger than the literature values. In particular, the $W=r-2.796(g-r)$ PWZ relation, which has the smallest metallicity term, provides $\mu_W$ that are in better agreement with the literature values (see Figure \ref{fig_muw}), and therefore can be recommended for the future distance scale applications. On the other hand, applications of multiband PLZ relations tend to give a larger $\mu_\lambda$ than the literature values. We suspect this may be caused by inclusion of the evolved models when calibrating the theoretical PLZ (and PWZ) relations. This is because the input luminosities for the evolved models are 0.2~dex higher than the ZAHB models \citep{marconi2022}. Thus the corresponding absolute magnitudes are brighter and the derived $\mu$ tend to be larger. We recall that the comparisons of the observed amplitudes with the theoretical Bailey diagram (i.e. Figure \ref{fig_ampvsx}) also reveal that the evolved models are totally inconsistent with the observations.

In summary, based on the analysis of DP1 light curves of the known RR Lyrae, together with the template light curves \citep{braga2024} and theoretical relations \citep{marconi2022}, we suggest utilizing the well sampled RR Lyrae light curves available from future LSST data releases for various astrophysical applications. This is because the sparsely sampled light curves tend to be problematic and should be treated with caution. In addition, such analysis also emphasizes the need for the theoretical models, especially for the PWZ and PLZ relations, to be refined or improved in the future works related to Rubin-LSST data. Finally, a further rigorous test on various theoretical relations presented in \citet{marconi2022} or future theoretical improvements could be done on RR Lyrae in monometallic globular clusters, which are expected to have homogeneous metallicity and a well-constraint evolutionary status, and RR Lyrae in a given globular cluster should have the same distance.

\begin{acknowledgments}

  CCN acknowledges the funding from the National Science and Technology Council (NSTC, Taiwan) under the grant 114-2112-M-008-011. TARA is supported by the NSTC grant 113-2740-M-008-005. AB acknowledges the funding from the Anusandhan National Research Foundation (ANRF) under the Prime Minister Early Career Research Grant scheme (ANRF/ECRG/2024/000675/PMS). This research was supported by the International Space Science Institute (ISSI) in Bern/Beijing through ISSI/ISSI-BJ International Team project ID $\#$24-603 – ``EXPANDING Universe'' (EXploiting Precision AstroNomical Distance INdicators in the Gaia Universe). We thank the useful discussions and comments from an anonymous referee to improve the manuscript.
  
  This publication is based in part on proprietary Rubin Observatory Legacy Survey of Space and Time (LSST) data, and was prepared in accordance with the Rubin Observatory data rights and access policies. All authors of this publication meet the requirements for co-authorship of proprietary LSST data

  This material is based upon work supported in part by the National Science Foundation through Cooperative Agreements AST-1258333 and AST-2241526 and Cooperative Support Agreements AST-1202910 and 2211468 managed by the Association of Universities for Research in Astronomy (AURA), and the Department of Energy under Contract No. DE-AC02-76SF00515 with the SLAC National Accelerator Laboratory managed by Stanford University. Additional Rubin Observatory funding comes from private donations, grants to universities, and in-kind support from LSST-DA Institutional Members. This research uses services or data provided by the Rubin Science Platform at NSF-DOE Vera C. Rubin Observatory, which is jointly funded by the U.S. National Science Foundation and the U.S. Department of Energy, Office of Science.

    This research has made use of the SIMBAD database and the VizieR catalogue access tool, operated at CDS, Strasbourg, France. This research has made use of the International Variable Star Index (VSX) database, operated at AAVSO, Cambridge, Massachusetts, USA. This research made use of Astropy,\footnote{\url{http://www.astropy.org}} a community-developed core Python package for Astronomy \citep{2013A&A...558A..33A,2018AJ....156..123A,2022ApJ...935..167A}.

  This work has made use of data from the European Space Agency (ESA) mission {\it Gaia} (\url{https://www.cosmos.esa.int/gaia}), processed by the {\it Gaia} Data Processing and Analysis Consortium (DPAC, \url{https://www.cosmos.esa.int/web/gaia/dpac/consortium}). Funding for the DPAC has been provided by national institutions, in particular the institutions participating in the {\it Gaia} Multilateral Agreement.

\end{acknowledgments}



%

\facilities{Rubin:Simonyi (LSSTComCam), Rubin:USDAC}

\software{LSDB \citep{lsdb2025}, {\tt dustmaps} \citep{green2018}, {\tt apply\_ugrizy\_templates.py} \citep{braga2024}}


\bibliography{DP1_knownRRL}{}
\bibliographystyle{aasjournal}



\end{document}